\documentclass[manuscript]{acmart}

\AtBeginDocument{%
  \providecommand\BibTeX{{%
    \normalfont B\kern-0.5em{\scshape i\kern-0.25em b}\kern-0.8em\TeX}}}

\usepackage{longtable}
\usepackage{subcaption}
\usepackage{graphicx}
\usepackage{enumitem}
\usepackage{hyperref}

\copyrightyear{2024}
\acmYear{2024}
\setcopyright{rightsretained}
\acmConference[CHI '24]{Proceedings of the CHI Conference on Human Factors in Computing Systems}{May 11--16, 2024}{Honolulu, HI, USA}
\acmBooktitle{Proceedings of the CHI Conference on Human Factors in Computing Systems (CHI '24), May 11--16, 2024, Honolulu, HI, USA}
\acmDOI{10.1145/3613904.3642398}
\acmISBN{979-8-4007-0330-0/24/05}

\begin{document}
\title[A Scoping Study of Evaluation Practices for Responsible AI Tools]{A Scoping Study of Evaluation Practices for Responsible AI Tools: Steps Towards Effectiveness Evaluations} 
\author{Glen Berman}
\affiliation{%
  \institution{Australian National University}
  \country{Australia}
  }
\email{glen.berman@anu.edu.au}

\author{Nitesh Goyal}
\affiliation{%
  \institution{Google Research, Google}
  \country{USA}
  }
\email{teshgoyal@acm.org}

\author{Michael Madaio}
\affiliation{%
  \institution{Google Research, Google}
  \country{USA}
  }
\email{madaiom@google.com}

\renewcommand{\shortauthors}{Berman, et al.}

\begin{abstract}
Responsible design of AI systems is a shared goal across HCI and AI communities. Responsible AI (RAI) tools have been developed to support practitioners to identify, assess, and mitigate ethical issues during AI development. These tools take many forms (e.g., design playbooks, software toolkits, documentation protocols). However, research suggests that use of RAI tools is shaped by organizational contexts, raising questions about how effective such tools are in practice. To better understand how RAI tools are---and might be---evaluated, we conducted a qualitative analysis of $37$ publications that discuss evaluations of RAI tools. We find that most evaluations focus on \textit{usability}, while questions of tools' \textit{effectiveness} in changing AI development are sidelined. While usability evaluations are an important approach to evaluate RAI tools, we draw on evaluation approaches from other fields to highlight developer- and community-level steps to support evaluations of RAI tools' effectiveness in shaping AI development practices and outcomes.
\end{abstract}




\maketitle

\section{Introduction}
\label{sec:intro}

Researchers concerned about issues of fairness, accountability, transparency, and equity in algorithmic systems are increasingly focusing on artificial intelligence (AI) development practices as a site for intervention \cite[e.g.,][]{Raji2021-gl,Madaio2020-ou,Rakova2021-mc,Holstein2019-lq}. In particular, the potential for AI systems\footnote{We follow the United States' National Institute for Standards and Technology definition of \textit{AI system}, which posits that an AI system ``is an engineered or machine-based system that can, for a given set of objectives, generate outputs such as predictions, recommendations, or decisions influencing real or virtual environments'' \cite[p.1]{tabassi2023artificial}.} to discriminate and exacerbate societal inequities \cite{Shelby2022-hr,buolamwini2018gender,obermeyer2019dissecting,koenecke2020racial} has prompted the interdisciplinary responsible AI (RAI) community to develop frameworks, resources, and processes for proactively identifying and mitigating potential societal impacts throughout system development and deployment \cite{Mittelstadt2019-ia, Madaio2020-ou, barocas2021designing, Holstein2019-lq, Morley2020-yp,Wong2022-zi}. As part of this effort, a range of tools have been developed for use by AI practitioners \cite[e.g.,][]{Lee2021-fn, Madaio2020-ou, Raji2020-wh}, stakeholders \cite[e.g.,][]{Selbst2021-ig,Reisman2018-df, Sandvig2014-wg,Raji2020-wh,devos2022toward, lam2022end}, and community members \cite[e.g.,][]{Krafft2021-no, shen2021everyday, Cooper2022-vz, coston2023validity}, to support RAI at all stages of the AI development and deployment pipeline.

Responsible AI tools are designed to influence the norms, decisions, and daily practices of AI development to mitigate the potential for AI systems to introduce or exacerbate societal inequities---RAI tools are designed not only to be widely used, but to be widely impactful \cite{Morley2020-yp, Mittelstadt2019-ia}. As such, the question of what it means for an RAI tool to be effective is important for the developers of RAI tools, their intended users, and the broader field. Further, as RAI tools are increasingly relied upon by technology firms to support the translation of ethical AI principles into practices \cite[e.g.,][]{Morley2023-rp}, their effectiveness is also of relevance to policymakers developing policies or regulation for AI systems \cite[e.g.,][]{tabassi2023artificial} and to communities who may be impacted by the deployment of such systems. Yet, while RAI tools have been studied from the perspective of their coverage of the ML pipeline \cite{Morley2020-yp}, their use by data scientists \cite[e.g.,][]{Lee2021-fn, deng2022understanding,Richardson2021-kp}, their suitability for specific tasks or contexts \cite{Lee2021-fn, Rostamzadeh2022-ie, sambasivan2021re}, and the normative commitments implicit in their design \cite{Wong2022-zi, Petterson2023-jc}, they have largely not been considered from the perspective of how they are (or might be) evaluated. As such, in this paper we ask: \textit{what are existing practices for evaluating RAI tools, as reported in publicly-available documentation?}
 
We address this research question through a thematic analysis of $37$ publications, within which evaluations of $27$ RAI tools are discussed. We focus on publicly-available publications as only publicly-available documents (rather than, e.g., internal, unpublished evaluations) are available to inform decisions about which RAI tools to adopt or recommend. We find that evaluations of RAI tools largely focus on questions of usability, primarily studied through small-scale qualitative evaluations conducted in in-lab settings. We demonstrate that such evaluations, while a valuable aspect of RAI tool development, are distinct from determining whether an RAI tool is \textit{effective} in achieving its desired aims. That is, most evaluations do not consider whether the overarching objective informing the design of an RAI tool---often expressed in terms of desired changes to AI development and deployment practices---is achieved. In this context, \textit{effectiveness} refers to the extent that a causal relationship can be established between a desired outcome and the use of an RAI tool in a given context \cite{cartwright2009}.\footnote{See Section \ref{relatedwork:evaluation_goals} for a discussion of effectiveness and the related concept of usability, and Section \ref{relatedwork:evaluation_validity} for a discussion of effectiveness and efficacy.} We extend our analysis by considering best practices in evaluations from the field of education---a field in which the evaluation of the effectiveness of interventions in educational processes has received significant research and public policy attention. Based on these best practices, we outline desiderata for the development of an \textit{effectiveness evaluation framework} and suggest practical steps for both RAI tool developers and the RAI and HCI communities to take towards developing and implementing such a framework.

This paper helps to bridge the gap between field-level overviews of RAI tools \cite[e.g.,][]{Morley2020-yp, Wong2022-zi} and empirical studies of RAI tool usage \cite[e.g.,][]{Lee2021-fn, Richardson2021-kp, Deng2022-rj,Madaio2022-db, Kaur2020-ht}. The paper contributes a critical perspective on RAI tool evaluation practices (and their gaps) through an analysis of RAI tool publications, and recommendations for the HCI and RAI communities to support effectiveness evaluations of RAI tools. Across these two contributions, we emphasize the significance of action at the level of the RAI and HCI communities, as RAI research and implementation in industry practice faces substantial resource constraints \cite[e.g.,][]{Holstein2019-lq,Madaio2022-db,Rakova2021-mc}. We call for community-level interventions to address these constraints and improve opportunities for more robust evaluation of RAI tools.

\section{Related work}
\label{sec:related_work}

In this section we provide an overview of prior work developing and studying RAI tools. We then introduce key concepts in evaluation, to contextualize our discussion of evaluation practices of RAI tools. 

\subsection{Defining and developing RAI tools}
\label{relatedwork:RAI_interventions}

We define \textit{RAI tools} as interventions that support AI practitioners and other stakeholders in addressing ethical issues in AI system development and deployment.\footnote{We note that there is ongoing debate regarding the scope and allocation of ethical responsibility during AI development and deployment \cite[e.g.,][]{Raji2022-by,Moss2020-bz, widder2022dislocated, Oduro2022-nz}. It is outside the scope of this paper to provide an overview of ethical issues in AI development \cite[cf.][]{jobin2019global, Morley2020-yp}. For our purposes, as discussed in Section \ref{methods:identifying_evals}, we adopt an inductive approach: we focus on tools which have been described by their creators as addressing ethical issues. We note, also, that creators may use a variety of terms besides `tool' (e.g., `framework', `toolkit', `software') to describe their work.} Reflecting the interdisciplinary nature of AI research and development, researchers and practitioners developing RAI tools have drawn from, and contributed to, a wide range of fields, including HCI \cite[e.g.,][]{coston2023validity, Crisan2022-zz, Vorvoreanu2023-im}, design \cite[e.g.,][]{IDEO_compass-bg, Boyd2022-wj}, software engineering \cite[e.g.,][]{Agarwal2018-ka, hutchinson2021towards}, and public policy \cite[e.g.,][]{Contractor2022-he, Reisman2018-df, AI_Now_Institute2018-hu}. For reviews of RAI tools, see \citet{Morley2020-yp,Petterson2023-jc}, and \citet{Wong2022-zi}.

Following Morley et al. \cite{Morley2020-yp} and Black et al. \cite{black2023operationalizing}, the breadth of RAI tools developed is reflected in their coverage across all stages of the AI development and deployment pipeline. RAI tools include: frameworks for guiding problem formulation and decisions to procure an AI system for a given use case \cite[e.g.,][]{IDEO_compass-bg, coston2023validity}; guides for considering ethical issues when designing an AI system \cite[e.g.,][]{Vorvoreanu2023-im, Boyd2022-wj}, such as design playbooks \cite{Hong2021-vj}, checklists \cite{DrivenData2019, Madaio2020-ou}, and workshop guides to support product teams to operationalize ethical guidelines in AI development \cite{Ballard2019-nw, Avin2020-lo}; approaches for enabling community participation in the development process \cite[e.g.,][]{Birhane2022-gc, Shen2022-bl,delgado2023participatory}; protocols and templates for curation, evaluation, and documentation of training datasets \cite[e.g.,][]{Bender2018-he, Gebru2021-ot, Fabris2022-ba, diaz2022crowdworksheets,denton2020bringing,davani2022dealing}; software tools for conducting fairness assessments of trained models \cite[e.g.,][]{Arnold2019-kb,Agarwal2018-ka, Bellamy2019-ho, Tenney2020-fo,Bird2020-lu,weerts2023fairlearn, barocas2021designing, Madaio2022-db}; protocols and templates for documentation of trained models and their evaluations \cite{Mitchell2019-aq,hutchinson2021towards}; and, processes for auditing AI systems by third parties \cite[e.g.,][]{Krafft2021-no,Selbst2021-ig,Reisman2018-df, Sandvig2014-wg,Raji2020-wh,devos2022toward,shen2021everyday,lam2022end, Metcalf2021-fd}. Also reflecting the breadth of RAI tool development is the range of intended users. RAI tools have been developed for use by AI practitioners \cite{Lee2021-fn, Madaio2020-ou, Raji2020-wh}, stakeholders \cite[e.g.,][]{Selbst2021-ig,Reisman2018-df, Sandvig2014-wg,Raji2020-wh,devos2022toward, lam2022end}, and community members \cite[e.g.,][]{Krafft2021-no, shen2021everyday, Cooper2022-vz, coston2023validity}.

We situate RAI tools within the space between high-level ethical AI principles \cite{data_framework_2018-py, Van_de_Poel2016-gf, jobin2019global} and low-level formalizations of intrinsically social and unobservable concepts, such as fairness metrics \cite[e.g.,][]{Puranik2022-bn, Fish2016-nt, corbett2018measure,dwork2012fairness}. High-level principles help shape the broader environment in which practitioners work, but are generally too abstract to be directly applied by practitioners \cite{Holstein2019-lq}. Similarly, low-level mathematical formalisms of social phenomena like fairness \cite{corbett2018measure} and algorithms that operationalize the measurement of these concepts are critical (and often contested \cite{jacobs2020meaning}) inputs into RAI tools, but without integration into a tool, they require a substantial level of technical expertise to put into use during AI development.

\subsection{Evaluating RAI tools}

The development of RAI tools is a significant undertaking. As researchers who have developed RAI tools (see our positionality statement in Section \ref{3.3}), we are familiar with the resource constraints that must be overcome to develop an RAI tool and convince practitioners to adopt it. Within this context, we recognize that a narrow focus on evaluation brings risk, including suppressing innovation \cite{greenberg2008usability}. Our view, however, is that greater focus on evaluation may be a path towards greater impact of RAI tools. By robustly substantiating the effectiveness of RAI tools, we can make a more compelling argument for their adoption (as \citet{Rakova2021-mc} explore for RAI work more generally). Yet, while the translation of ethical AI principles into industry practices and tools has received attention in HCI and adjacent fields \cite[e.g.,][]{varanasi2023it, wang2023designing, Holstein2019-lq,Madaio2020-ou,Rakova2021-mc, Vorvoreanu2023-im, elsayed-ali2023responsible, Lee2021-fn}, relatively little attention has been directed to evaluating RAI tools' \textit{effectiveness}. Recent surveys of RAI tools \cite{Lu2022-ru, Lu2022-rq, chen2022, Morley2020-yp}, and studies of particular types of tools \cite{chen2022} highlight the challenges of designing effective RAI tools, and of integrating RAI tools into the complexities of an AI development pipeline \cite{black2023operationalizing,Deng2022-rj,Holstein2019-lq,Madaio2020-ou, Madaio2022-db, Rakova2021-mc,Wong2022-zi}.

Questions of tool usability and usefulness have been investigated in recent studies of RAI tools: six open source tools for fairness testing were compared in an interview and survey study \cite{Lee2021-fn}, and the usability of two of these tools was studied in a think-aloud evaluation with practitioners encountering the tools for the first time \cite{Deng2022-rj}. Similarly, two software tools for fairness testing were assessed through a simulated scenario with practitioners \cite{Richardson2021-kp}, and RAI tools to support AI system auditing and impact assessments were mapped against standards for audits and impact assessments in other domains \cite{Ayling2022-wm,moss2021assembling,Metcalf2021-fd,Selbst2021-ig}. These evaluations provide valuable insights on the usability challenges facing RAI tool developers, which are critical to address in order to improve the likelihood of tool adoption.

As RAI tool development continues to mature, enhancing and expanding usability evaluation practices will be important. As usability studies unblock adoption, we argue that evaluations that aim to address the \textit{effectiveness} of an intervention will be necessary and complementary extensions to usability evaluations. \textit{Effectiveness} is the extent to which a causal relationship can be established between an intervention and a desired outcome in a given context \cite{cartwright2009}. In contrast, \textit{usability} in an evaluation context refers to the ease with which particular groups can implement the tool or intervention \cite{Ledo2018-yt}.\footnote{See \cite{Bevan2001-li} for an overview of usability standards and definitions in the HCI field.} Effectiveness and usability are thus contextually interdependent concepts, with their operationalization varying from one intervention to another. As one example, a dataset documentation protocol may aim to increase transparency in the ML pipeline, reduce the likelihood of data misuse, and improve accountability for dataset creation and model training decisions \cite[p.1]{Fabris2022-ba} . The effectiveness of a dataset documentation tool against each of these aims is a reflection of the causal relationship between adoption and use of the protocol in a particular context (e.g., a given technology firm) and measurable changes in system development practices or outcomes (e.g., reduction in data misuse); although, as we discuss in Section \ref{improving:validity}, such measures may not currently exist for many responsible AI goals, or may be difficult to develop due to contestation in operationalizing constructs such as fairness \cite{Jacobs2021-bk}.

Prior work demonstrates the importance of questions of effectiveness, and highlights the risk of failing to attend to the evaluation of RAI tools. Studies have found that organizational cultures, structures, and incentives can act as barriers to the effective operationalization of ethical AI principles \cite{Madaio2020-ou,Schiff2020-rw,Rakova2021-mc}. Chang and Custis \cite{Chang2022-jc} find that organizational structures and incentives can present implementation challenges for the effective use of documentation templates. Metcalf et al. \cite{Metcalf2021-fd}, drawing from fields adjacent to ML, found that algorithmic impact assessments, intended to highlight risks of AI system deployment, can instead be co-opted by firms developing such systems to further their interests. Meanwhile, recent work has also explored and critiqued evaluation practices of AI systems more broadly, highlighting the implications of decontextualization when evaluating AI systems \cite{Hutchinson2022-gh, mehandru2022reliable} and raising attention to the risks of corporate capture when private actors evaluate their own systems \cite{Young2022-pi}. Our aim is to support improvements in the design and development of RAI tools through an analysis of existing evaluation practices for RAI tools.

\subsection{Tool evaluations in HCI}

While there have been advances in the field of RAI tool development, evaluation of RAI tools is still a nascent area of research and practice. In contrast, tool evaluation more generally is a longstanding focus of research and practice within HCI \cite{myers1996strategic}. In HCI research, evaluations have historically tended to focus on questions of tool usability, which are frequently investigated through user studies \cite{barkhuus2007mice, greenberg2008usability}. However, a focus solely on usability evaluations has been critiqued across HCI research areas, notably by \citet{olsen2007evaluating} and \citet{greenberg2008usability}.

\citet{olsen2007evaluating}, in his analysis of user interface systems research, highlights the `usability trap'. On the one hand, the assumptions and practical constraints of usability experiments are difficult to reconcile with the complexity of user interface systems. On the other, running a usability experiment, however flawed, is perceived as a requirement for publishing. \citet{olsen2007evaluating} argues that evaluation methods should be grounded in the claims researchers hope to make, and proposes a framework of "Situations", "Tasks", and "Users" for thinking about the kinds of claims one might make, and therefore the sorts of evaluation activities that may be necessary. Greenberg and Buxton \cite{greenberg2008usability} similarly argue that evaluation design decisions must flow from the research question(s) under consideration, and are also critical of the predominance of usability evaluations in HCI research. They note that the kinds of claims one might make also differ according to the design status of the system under evaluation: early design sketches intended to demonstrate a novel approach need not be evaluated in the same way as working prototypes. Lastly, Greenberg and Buxton \cite{greenberg2008usability} also note that usability and usefulness are distinct concepts: a tool can be usable but useless.

Of particular relevance to our focus are evaluations of interactive design toolkits \cite{Ledo2018-yt, myers2000present}, which, like many RAI tools, are designed to be used by practitioners to support the development of computing artefacts (here, developing software; for RAI tools, developing AI models \cite{Holstein2019-lq} or AI-powered applications \cite{wang2023designing}). \citet{Ledo2018-yt} reviewed publications from leading HCI venues in which interactive design toolkits are discussed, and identified four types of evaluations: demonstrations, which show what the toolkit might enable; usage, which use qualitative evaluations to show the usability of the toolkit; technical, which use software engineering techniques to evaluate technical performance; and heuristics, which are informal assessments of toolkit usability through the application of design guidelines. Echoing others \cite[e.g.,][]{greenberg2008usability, olsen2007evaluating}, \citet{Ledo2018-yt} reiterate that evaluation methods for a specific toolkit should be determined in light of the claims researchers wish to make about the toolkit.\footnote{A similar emphasis on alignment of evaluation design and research claims can be found in software engineering literature on user evaluations. We refer interested readers to \citet{ko2015practical}.}\looseness=-1

As we explore further in Sections \ref{improving:validity} and \ref{improving:effectiveness_framework_field_changes}, we agree with the call to align evaluation methods with research claims, and draw on HCI guidelines for evaluation design that have been developed in response to this call \cite[e.g.,][]{myers2000present, davis2023what} to inform an RAI tool effectiveness evaluation framework. As RAI tools aim to address ethical issues in AI development, claims related to the consequences of their use or effectiveness are wide in their sociopolitical scope, implicating not only RAI tool developers and users, but also AI system stakeholders and communities who may be affected by AI system deployment. This necessitates evaluation designs of commensurate scope, and prompts us to consider how fields outside of HCI conceptualise and design evaluations.

\subsection{Evaluation goals and approaches outside of HCI}
\label{relatedwork:evaluation_goals}

The evaluation of trained models, and of automated systems incorporating these models, is a significant area of research in the ML community. In part, these research efforts are responses to reproducibility concerns in ML, which have prompted researchers to consider how claims about model performance can be better evaluated \cite{hutson2018artificial, pineau2021improving, Liao2022-fu}. These efforts also respond to critical analyses that demonstrate that evaluation of models, in isolation from the systems they are deployed within, is incomplete---model performance on a benchmark dataset may not be predictive of system behaviour \cite{Raji2020-wh, Raji2022-by}. In natural language processing (NLP), for instance, researchers have surveyed existing NLP model evaluation methods, finding no standardised evaluation practices\cite{gehrmann2023repairinga, zhou2022deconstructing}. Relatedly, efforts are underway to develop standards for evaluation of ML applications \cite{hammond2021framework} and models \cite{Hutchinson2022-gh}. Our focus complements these efforts, by attending to the evaluation of interventions in ML production, rather than the evaluation of the outputs of ML production, such as trained models or new AI systems.

In doing so, we look to education \cite{towne2002scientific} and medicine \cite{Craig2008-of}, two fields outside of ML and HCI, which have also suffered from reproducibility concerns \cite[e.g., in education research,][]{schneider2004building}, and have developed robust norms for the evaluation of process interventions (e.g., implementing a new curriculum), despite the challenges of dynamic and heterogeneous intervention contexts like schools and hospitals. In education, for example, decades of research and public policy have identified the need to develop standards and norms for evaluating the effectiveness of an intervention. Government agencies have sought evidence to inform investment in interventions such as new curricula or learning technologies \cite{Alkin2004-px}. In response, the U.S.\ Department of Education’s Institute for Education Sciences (IES) created a ``What Works Clearinghouse'' to provide a standard process and repository for evaluating the evaluations of educational interventions, to inform teachers, curriculum designers, school leaders, and policymakers about what works in improving educational outcomes \cite{clearinghouse2012works, cushing2023challenging}.

The fields of education and medicine are imperfect analogues for RAI. We can, however, draw lessons from their approaches to evaluation. In doing so, we acknowledge that evaluations practices in these fields are imperfect, and continue to be debated. Education and medicine interventions aim to be highly generalisable, but have to contend with varied social contexts and the numerous confounding variables these contexts present (e.g., see discussion of confounding variables in education interventions in the What Works Clearinghouse procedure manual \cite{whatworks2022} and discussion of different social contexts for medical interventions in \cite{cartwright2010limitations}). In both fields, interventions are a site of policy debate and regulatory action \cite[e.g.,][]{ghate2001community}. Similarly, many RAI tools aim to be generalisable to a wide range of contexts, such as different types of technology companies, different application domains for which AI systems are developed, different cultures or geographies of deployment contexts of a given AI system, as well as relevance for various stages of the AI development pipeline.\footnote{We note there are some efforts to develop RAI tools for specific application domains, such as health \cite[e.g.,][]{Rostamzadeh2022-ie}, or RAI frameworks for different cultural contexts \cite[e.g.,][]{sambasivan2021re}, which we discuss further in Section \ref{improving:validity}.} And, as efforts to enhance regulatory oversight of AI mature, one area of focus is on best practices for AI production pipelines \cite[e.g., in a health context,][]{kim2023organizational}, which may extend to use of RAI tools. Yet, unlike the education and medicine fields, these RAI efforts are occurring against a historical backdrop of minimal public oversight (despite some emerging AI risk frameworks from government agencies in some countries \cite{tabassi2023artificial}), and scant formal accreditation for AI practitioners.

It is important to note that different types of evaluations have different goals. For instance, \citet{gertler2016impact} divide intervention evaluations into several types: descriptive evaluations that seek to answer questions about the current state of the world (e.g., understanding processes, conditions, relationships), normative evaluations that compare what is taking place with what should take place (e.g., with what fidelity did teachers implement a particular curricular intervention), and causal evaluations, which explore the effect a particular intervention has on behaviors, attitudes, or other outcomes of interest. Different evaluation goals entail different types of studies, including observational (which may involve case studies, surveys, etc.), design-based research \cite{barab2004design, nathan2010learning}, quasi-experimental (e.g., difference-in-difference), and experimental studies (e.g., randomized controlled trials). In many fields, experimental studies are held up as the ``gold standard'' for evaluation quality (although not without critiques \cite[e.g.,][]{connolly2018trials,styles2018randomised,morrison2001randomised}), as they allow researchers to compare the outcomes (e.g., test scores) of a group who received an intervention with a similar control group that did not \cite{gertler2016impact, clearinghouse2012works}. When randomized assignment of participants to a treatment or control group makes experimental studies impossible or infeasible, as in state-level health policy research \cite[e.g.,][]{abouk2021immediate}), researchers may use quasi-experimental methods (e.g., regression discontinuity analysis \cite{post2021impact}, difference-in-difference \cite{abouk2021immediate}, propensity score matching, and instrumental variables \cite{hedberg2017body,higgins2011racial}) to approximate the controls used in an experimental study.

A longer discussion of these evaluation methods is beyond the scope of this paper.\footnote{We refer readers to \cite{towne2002scientific,gertler2016impact,clearinghouse2012works} for more details.} However, we note that each method presents its own set of compromises and trade-offs, which need to be considered before adapting them to RAI contexts, as we discuss further in Section \ref{improving:effectiveness_framework_field_changes}. Additionally, education and medicine have also suffered from reproducibility crises \cite[e.g., in education research,][]{schneider2004building}, with their evaluation methods continuing to develop in response. Of particular relevance are differences in the way RAI tools are developed and deployed compared to interventions in the education or medicine fields. As outlined in Section \ref{relatedwork:RAI_interventions}, many, but not all, RAI tools are software tools, often developed within an agile paradigm, or are open-source software where the functionality may change over time \cite[e.g.,][]{weerts2023fairlearn}. In this context, implementing experimental studies, which assume that the intervention will remain static, may be challenging. In addition, many experimental studies assume a controlled intervention environment, in which confounding variables are well understood \cite{cartwright2010limitations, cowen2017randomized}. Yet many RAI tools are not only designed within an agile paradigm, but are also intended to be used in an agile AI development paradigm, where work practices are continually changing.

In our view, however, these challenges should not forestall consideration of how evaluation methods from other fields can be adapted for use in RAI tool evaluations. Indeed, research in educational contexts like schools is also subject to changing policies, state standards, students moving from an intervention to control group when enrolling in a new class or school, as well as exogenous shocks like teacher strikes, political violence, and school closures, among others \cite{kizilcec2021mobile}. Conversely, as AI technologies and development paradigms are increasingly incorporated into medical or education interventions, these domains are also starting to consider how to apply their familiar evaluation approaches to AI technologies \cite[e.g., in healthcare,][]{reddy2021evaluation, magrabi2019artificial}\cite[e.g., in education,][]{cardona2023artificial}, and how to translate insights from RAI to their evaluation practices \cite[e.g.,][]{siala2022shifting}. These efforts demonstrate the potential of methods drawn from otherwise disparate disciplines for guiding new evaluation practices, although doing so successfully will require a thoughtful approach to managing threats of evaluation validity.

\subsection{Evaluation validity}
\label{relatedwork:evaluation_validity}

Regardless of the goal of the evaluation, there are criteria for assessing the validity of the evaluation.\footnote{For an example of how these criteria apply to NLP model evaluations, see \citet{Liao2022-fu}.} For studies that attempt to establish a causal relationship, \textit{internal validity} is one way of understanding whether a given intervention causes the intended effect, and refers to the aspects of a study design that ``support causal claims between variables'' \cite{flake2020measurement}.\footnote{A related concept, \emph{construct validity}, has been explored by \citet{Jacobs2021-bk} in relation to algorithmic fairness, discussing how operationalizing an unobservable phenomena (e.g., fairness) in terms of observable measurables such as  fairness metrics is a process of measurement modeling, and should be interrogated and documented to provide evidence that measures are capturing what they are intended to measure.} For experimental evaluations, this might involve exploring threats to internal validity, such as alternative causes other than the intervention, study attrition, or leakage between a treatment and control group (e.g., if a student in the control group class switches into a class receiving the treatment) \cite{clearinghouse2012works}.\looseness=-1

In addition, \textit{external validity} is the aspect of a given study that captures the extent to which the findings from that study are able to be generalized to additional populations or contexts (e.g., new study sites, or other geographic or cultural contexts) \cite{nathan2010learning, gertler2016impact, bold2013scaling}. In other words, how do you know whether your findings will be relevant for others who did not participate in the study? Some ways to avoid threats to external validity include conducting multi-site studies, random sampling, evaluating the representativeness of the sample to the population as a whole,\footnote{See \citet{chasalow2021representativeness} for a longer discussion of the slippery meanings of the term `representativeness' as it is used across different fields.} and more broadly, being transparent in documentation of the study site, population, recruitment and inclusion criteria, and other methodological decisions \cite{flake2020measurement}.

One key component of external validity is the \textit{ecological validity} of the study, or the extent to which the study is reflective of a ``real world’’\footnote{See \ref{improving:validity} for a discussion of how this term may lead to homogeneity in thinking about the range of contexts where ML systems are developed and used.} \cite{Holleman2020-vl}. Originating in critiques of in-lab psychology studies that were not reflective of people’s behavior or contexts outside of the lab, the term, despite its popularity, has suffered from ambiguous usage and is too often used as a shorthand when discussing the specific contexts in which a study is conducted (and for which the results might be more broadly applicable) \cite{Holleman2020-vl}.

An additional related concept is \textit{efficacy}, the extent to which an intervention produces a desired outcome in ideal experimental settings \cite{cartwright2009}. That desired outcome might be expressed in terms of usability, or effectiveness, or some other goal. But, crucially, the extent to which an efficacy evaluation is valid outside of the in-lab settings under which it is conducted---i.e., the extent to which it has external validity---depends on the desired outcome under consideration. Consider an evaluation design that consists of an observational study, in which participants will follow an interactive tutorial to use a software toolkit to explore a dataset. Threats to the external validity of this evaluation vary depending on whether the objective is to test the usability of the software toolkit or its effectiveness. As an evaluation focused on usability, one might say that the in-lab settings closely approximates the context in which the software toolkit is intended to be used, e.g., because participants use a computer that is similar to their workplace computer, and a dataset that they also use in their daily work. As such, one might say that confirming the efficacy of the software toolkit through this in-lab evaluation is likely to indicate that intended users will be able to successfully use the software toolkit. However, if the aim of the evaluation is to determine the toolkit's effectiveness, then the limitations of this efficacy evaluation design are much more acute. The in-lab settings have only considered interactions between individual users and the software toolkit, not the relationship between an individual user (using the software toolkit) and the broader collaborative process of developing an AI system. As such, there is a misalignment between this efficacy evaluation design and the objective of an effectiveness evaluation. In short, an efficacy evaluation asks ``can the RAI tool work?'', whereas an effectiveness evaluation asks ``when used, is the RAI tool impactful?'' \cite{Fedson1998-ej, Bevan2001-li}. It is the latter question which we argue is of most relevance to the evaluation of RAI tools. We return to this tension in more depth in Section \ref{statusquo:patterns_usability}.\looseness=-1

Finally, it is critical to acknowledge that there may be tradeoffs between the types of validity---for instance, a study that has high ecological validity (e.g., an \textit{in-situ} classroom study) may suffer from issues with internal validity (i.e., due to students changing classes or dropping out in the middle of the study); a laboratory study that provides compelling evidence of efficacy may only have external validity for a very narrow range of contexts. Although other fields have extensive literature documenting these potential threats to validity and their tradeoffs for particular evaluation goals and methods, the field of RAI has had little such investigation. In this paper, we intend to open this conversation.

\section{Methods}
\label{sec:methods}

The research question informing this study is: \textit{what are existing practices for evaluating RAI tools, as reported in publicly-available documentation?} To address this question, we analyze publicly available documentation of RAI tools. Inspired by similar approaches in CHI \cite{Petterson2023-jc, Ledo2018-yt} and CSCW \cite{Wong2022-zi}, we use publicly-available documentation produced by researchers and practitioners as a primary source through which to explore evaluation practices. While RAI tool evaluations may be documented in other ways (e.g., internal reports or monitoring dashboards), we limit our study to publicly-available documentation because, from the perspective of potential RAI tool users, policymakers, or other stakeholders, only publicly-available documentation can inform decisions about which RAI tools to deploy in a given context.\footnote{Indeed, in other fields, policymakers have required evaluations to be publicly documented, as discussed in Section \ref{relatedwork:evaluation_goals}.} We discuss the limitations of this approach in Section \ref{sec:limitations}.

\subsection{Developing the corpus of publications}
\label{methods:identifying_evals}

To the best of our knowledge, there is currently no central repository for evaluations of RAI tools, or indeed for RAI tools themselves (see Section \ref{relatedwork:RAI_interventions} for a description of existing tools and surveys of tools).\footnote{A repository of datasets used in studies of algorithmic fairness is under development \cite{Fabris2022-ba}.} Additionally, as existing surveys of RAI tools demonstrate, publication practices regarding RAI tools vary widely: RAI tools have been published in HCI and ML academic venues \cite[e.g.,][]{Vorvoreanu2023-im, elsayed-ali2023responsible, Luccioni2022-ow, Mitchell2019-aq}, on organizations' websites \cite[e.g.,][]{Reisman2018-df, Bird2020-lu}, and on  GitHub \cite[e.g.,][]{audit_ai-uh}. As such, following \cite{Wong2022-zi} and in keeping with the exploratory nature of this study, we adopted a purposive sampling strategy \cite{Etikan2016-ka, Wong2022-zi,Petterson2023-jc}, whereby our goal was to sample a variety of tool types, creators, and dimensions of responsible AI (e.g., fairness, transparency, accountability), rather than to develop an exhaustive or statistically representative sample.

We conducted our search for RAI tools between October and December, 2022. We created an initial list of RAI tools by mining reviews of RAI literature \cite{Morley2020-yp, Ayling2022-wm, chen2022, Wong2022-zi, Lee2021-fn}. We used this list to design a keyword search of the ACM Digital Library (DL),\footnote{The ACM DL search used the keywords `fair*', `ethic*', `responsib*', `bias', `discrimination', or `equity' occur alongside the keywords `AI', `artificial intelligence', `ML', or `machine learning' and the keywords `tool*' or `intervention', and filtered for research articles.} which includes SIGCHI publications and the proceedings of the AIES and FAccT conferences. One author reviewed the title and abstracts of publications returned by the ACM DL search to identify further RAI tools. This search process resulted in an initial list of $243$ unique RAI tools, with each tool linked to at least one publication about it. To select a subset for qualitative analysis we used a two-stage filtering process. First, one author applied the following exclusion criteria, designed to ensure alignment with our definition of RAI tools (see Section \ref{relatedwork:RAI_interventions}):
\begin{itemize}
    \item \textit{Unavailable}: $5$ tools were excluded because they are now obsolete or cannot be found online; $2$ were excluded because they are only available through commercial licenses.
    \item \textit{Category error}: $6$ potential tools were excluded as they were organizations that provide services, rather than tools; $14$ were excluded because they were information provision services (e.g., websites that track resources for journalists \cite{algo_tips-ql}), rather than tools; $15$ were excluded because they were conceptual contributions (e.g., a proposal to use the concept of a windfall tax as a tool for redistributing technology firm profits \cite{OKeefe2020-zh}), rather than tools.
    \item \textit{Low-level construct}: $35$ potential tools were excluded as they were low-level formalizations (e.g., the Shifted Decision Boundary method for optimizing the trade-off between bias and accuracy in a classifier \cite{Fish2016-nt}), which fall outside our definition of an RAI tool, as they are not directly usable by practitioners or other stakeholders.
    \item \textit{High-level framework}: $17$ potential tools were excluded because they were high-level ethical frameworks, principles, or guidelines (e.g., the Ontario Privacy Commissioner's Privacy by Design principles \cite{Cavoukian2010-lg}), which do not meet our definition of an RAI tool, as they require operationalization before they can be directly used by practitioners or other stakeholders.
    \item \textit{No relevance to RAI issues}: $4$ tools were excluded because their creators did not describe them as addressing issues of responsible AI (e.g., fairness, accountability, transparency, or equity). For example, Flipper \cite{Varma2017-tp}, a tool for debugging training sets, was excluded.
\end{itemize}
The application of these exclusion criteria left $145$ RAI tools and their associated publications. To reduce this to a manageable number of tools for qualitative analysis, following \citet{Wong2022-zi} and \citet{Petterson2023-jc}, we sampled for breadth, using the following inclusion criteria: stages of the pipeline of AI design, development, and deployment; the types of RAI tools proposed; the sectors of tool creators (i.e., academic, industry, civil society); and different forms of evidence for tool adoption (e.g, GitHub, paper citations, commercial platforms).\footnote{We attempted to sample tools that were widely adopted, but direct measures of adoption (e.g., monthly users for a software tool) are not publicly available; as such, we rely on proxies for adoption \cite{Wong2022-zi}, as shown in Table \ref{tab:use}. In the final sample, we were unable to find evidence of use for seven tools.} To apply these inclusion criteria, one author reviewed the full text and metadata of all publications and classified the publications accordingly. Two authors reviewed these classifications and together identified a sample corpus of publications with breadth across all criteria for thematic analysis. This corpus consisted of $27$ RAI tools (T1 - T27, listed in Appendix \ref{app:toollist}). These tools were associated with $37$ publications (A1 - A37, listed in Appendix \ref{app:publist}). The corpus is described further below in Section \ref{method:corpus}.\looseness=-1

\subsection{Data analysis}
\label{methods:analysis}

Then, we conducted an inductive thematic analysis \cite{Braun2006-ts, nowell2017thematic} on the corpus of $37$ publications, following \citet{Wong2022-zi} and \citet{Petterson2023-jc}. One author initially excerpted quotes related to toolkit evaluation (based on our research question) from each publication, which were discussed with another author. Then, through several rounds of inductive thematic analysis, using an online whiteboard, two authors iteratively generated codes to capture patterns of shared meaning across multiple quotes and clustered these codes into related themes, with all three authors discussing the final set of themes. A summary of all themes and codes, with example excerpts, is provided in Appendix \ref{app:thematicanalysis}.

\subsection{Corpus description}
\label{method:corpus}

\subsubsection{Review of RAI tools in the corpus}

As shown in Table \ref{tab:sectors}, the majority of RAI tools in our corpus are created within industry contexts (15), in addition to a mix of academic (6), civil society (3), and cross-sector creators (3). The plurality of RAI tools are software toolkits (10). Others are transparency artifacts (6), workshop guides or playbooks (5), third-party review tools (3) and more (see Table \ref{tab:types}). In Table \ref{tab:stages}, we categorise RAI tools in the corpus in terms of the stage of AI development for which they are designed to be used.\footnote{As some tools are under active development, their intended use may continue to evolve in the future.} We note that some tools can be used across multiple stages of development (e.g., during training, testing, and deployment), and AI development does not always move linearly through these stages. A description of the RAI tools categorised in each stage of AI development can be found in Appendix \ref{app:toolcats}.

\begin{table}[h!]
\footnotesize
    \centering
\caption{Evidence of use for RAI tools in the corpus.}
\label{tab:use}
\resizebox{\columnwidth}{!}{%
\begin{tabular}{@{}p{0.15\linewidth}p{0.17\linewidth}p{0.17\linewidth}p{0.17\linewidth}p{0.17\linewidth}p{0.17\linewidth}@{}}
\toprule
Evidence &
Citations (\textgreater{}20) & 
Github forks (\textgreater{}40) &
Platform integration & 
Industry adoption & 
Ongoing research outputs\\
\midrule
Tools & 
13 & 
8 & 
8 & 
8 & 
2 \\ 
\bottomrule
\end{tabular}%
}\end{table}
\begin{table}[h!]
\footnotesize
    \centering
\caption{Sectors of tool creators.}
\label{tab:sectors}
\resizebox{\columnwidth}{!}{%
\begin{tabular}{@{}p{0.2\linewidth}p{0.2\linewidth}p{0.2\linewidth}p{0.2\linewidth}p{0.2\linewidth}@{}}
\toprule
Sector & Academic & Industry & Civil society & Cross sector\\ \midrule
Tools & 6 & 15 & 3 & 3\\ 

\bottomrule
\end{tabular}%
}\end{table}
\begin{table}[h!]
\footnotesize
    \centering
\caption{Types of tools represented in the corpus.}
\label{tab:types}
\resizebox{\columnwidth}{!}{%
\begin{tabular}{@{}p{0.08\linewidth}p{0.1\linewidth}p{0.1\linewidth}p{0.1\linewidth}p{0.1\linewidth}p{0.1\linewidth}p{0.1\linewidth}p{0.1\linewidth}@{}}
\toprule
Tool type & 
Benchmarking platform & 
Transparency artifact & 
Dataset & 
Workshop guide &
License template & 
Third-party review & 
Software toolkit \\
\midrule
Tools & 
1 & 
6 & 
1 & 
5 & 
1 & 
3 & 
10 \\ 
\bottomrule
\end{tabular}%
}\end{table}

\begin{table}[h!]
\footnotesize
    \centering
\caption{Development stage targeted by tools in the corpus.}
\label{tab:stages}
\resizebox{\columnwidth}{!}{%
\begin{tabular}{@{}p{.08\linewidth}p{.08\linewidth}p{.08\linewidth}p{.08\linewidth}p{.08\linewidth}p{.08\linewidth}p{.08\linewidth}p{.08\linewidth}p{.08\linewidth}@{}}
\toprule
Stage & 
Problem formulation & 
Design phase &
Data collection \& processing & 
Training & 
Testing & 
Deployment &
Monitoring & 
Other \\ 
\midrule
Tools 
& 2 
& 2 
& 4 
& 2
& 9 
& 4 
& 3 
& 1 \\
\bottomrule
\end{tabular}%
}
\end{table}

\subsubsection{Review of publications in the corpus}
\label{method:review_pubs}

Publications in the corpus position their contributions in different ways: several publications describe their primary contribution as the evaluation of an RAI tool (e.g., [\hyperlink{A9}{A9}, \hyperlink{A24}{A24}, \hyperlink{A25}{A25}, \hyperlink{A29}{A29}]); in others, evaluation practices are discussed, but the primary stated contribution is a description or demonstration of an RAI tool (e.g., [\hyperlink{A2}{A2}, \hyperlink{A6}{A6}, \hyperlink{A8}{A8}, \hyperlink{A19}{A19}]). In several publications, evaluation practices are not explicitly discussed, although tools' capabilities or effectiveness are still considered. In three publications the tool developers report on their own experience of using the tool, which includes discussions of tool impact [\hyperlink{A4}{A4}, \hyperlink{A22}{A22}, \hyperlink{A32}{A32}]. In four publications a case study is provided to highlight how the tool has been used and to support claims related to effectiveness [\hyperlink{A2}{A2}, \hyperlink{A3}{A3}, \hyperlink{A16}{A16}, \hyperlink{A19}{A19}]. In four publications, worked examples are used to demonstrate the tool [\hyperlink{A1}{A1}, \hyperlink{A5}{A5}, \hyperlink{A6}{A6}, \hyperlink{A10}{A10}, \hyperlink{A14}{A14}]. In a small number of publications only technical descriptions of an RAI tool are provided [\hyperlink{A13}{A13}, \hyperlink{A17}{A17}, \hyperlink{A19}{A19}, \hyperlink{A26}{A26}]. Finally, several publications describe iterative cycles of development and evaluation of an RAI tool [\hyperlink{A7}{A7}, \hyperlink{A11}{A11}, \hyperlink{A18}{A18}, \hyperlink{A20}{A20}, \hyperlink{A34}{A34}, \hyperlink{A35}{A35}], highlighting close relationships between those developing RAI tools and those evaluating them. Similarly, several publications report the findings of evaluations conducted specifically to inform future tool development [\hyperlink{A9}{A9}, \hyperlink{A15}{A15}, \hyperlink{A23}{A23}, \hyperlink{A25}{A25}, \hyperlink{A31}{A31}].\looseness=-1

\subsection{Positionality statement}\label{3.3}

This paper is the product of the authors' shared interest in the development of RAI tools and AI development practices. All authors are male, based in the Global North, and work for an industry research institution in various capacities. Several authors have conducted prior research contributing to the development of AI tools, guides, and other resources, as well as research studying responsible AI development and evaluation practices. As authors based in industry, we are attuned to the challenges of development and use of RAI tools in applied industry settings, and thus we have primarily focused on how RAI tools can be better evaluated for their effectiveness in such settings (although we discuss implications for public policy). Future work should consider evaluation processes for the use of RAI tools in other settings, such as the public sector \cite[cf.][]{Veale2018-rs} or civil society.\looseness=-1

\section{RAI tool evaluation practices}
\label{sec:existing_practices}

In this section, we report the findings of our analysis of existing practices for evaluation of RAI tools. We structure this section by themes about existing practices. For each theme, we describe how it manifests in the corpus, and we discuss gaps between the existing practice and the theories of change for RAI tools. We emphasize that the gaps we identify are at the level of collective practice. No gap is uniformly present across all publications in our corpus. In some instances, the evaluation activity gaps we identify are also self-identified by authors in their evaluation write-up---our contribution is to demonstrate how these gaps manifest across a wide range of RAI tool evaluations. The themes we discuss include the focus of existing evaluation practices on usability (rather than effectiveness); evaluating fit with existing ways of working (rather than changing existing development practices); individual use (rather than use within teams or organizations); and, evaluating ``real world'' use (without specificity about what world(s) they are evaluating RAI tools for). In Section \ref{sec:improving_evaluation} we open a conversation about how the field might support the development and adoption of an effectiveness evaluation framework for RAI tools.

Table \ref{tab:eval_activities} provides an overview of evaluations reported in the corpus. The majority of RAI tools in the corpus were evaluated using qualitative methods drawn from HCI, including semi-structured interviews, think-aloud studies, and contextual inquiries \cite{olson2014ways}, as well as workshops or focus groups, and pilot or proof-of-concept case studies. For instance, contextual inquiry is used to explore whether users accurately interpret the visualizations produced by a machine learning interpretability tool [\hyperlink{A31}{A31}]. Think-aloud studies are used to explore the internal states of users as they interact with a software toolkit [\hyperlink{A20}{A20}, \hyperlink{A27}{A27}, \hyperlink{A30}{A30}] or transparency artifact [\hyperlink{A23}{A23}]. Interviews are used to probe whether users interact with a software toolkit as intended [\hyperlink{A15}{A15}, \hyperlink{A20}{A20}], and to identify challenges in adoption of transparency artifacts [\hyperlink{A7}{A7}, \hyperlink{A8}{A8}], a software toolkit [\hyperlink{A29}{A29}], and a workshop guide [\hyperlink{A33}{A33}]. Other mixed-methods approaches, such as surveys, are used in a smaller number of tool evaluations, for instance, as a data collection approach following a workshop or pilot study of a transparency artifact [\hyperlink{A7}{A7}], workshop guide [\hyperlink{A9}{A9}, \hyperlink{A21}{A21}], or software toolkit [\hyperlink{A20}{A20}], or surveys are used as a way to understand the broader applicability of findings arising from small-scale interview studies of software toolkits [\hyperlink{A27}{A27}, \hyperlink{A29}{A29}, \hyperlink{A31}{A31}].\looseness=-1

\begin{table}[t]
\centering
\caption{Evaluation activities discussed in the corpus.}
\label{tab:eval_activities}
\resizebox{\columnwidth}{!}{%
\begin{tabular}{@{}p{0.4\linewidth}p{0.1\linewidth}p{0.5\linewidth}@{}}
\toprule
Evaluation activity & Count & Publications \\ \midrule
Contextual inquiry & 2 & \hyperlink{A28}{A28}, \hyperlink{A31}{A31} \\
Focus group & 2 & \hyperlink{A7}{A7}, \hyperlink{A29}{A29} \\
Illustrative case study & 4 & \hyperlink{A2}{A2}, \hyperlink{A3}{A3}, \hyperlink{A16}{A16}, \hyperlink{A19}{A19} \\
Interviews (type unspecified) & 2 & \hyperlink{A7}{A7}, \hyperlink{A31}{A31} \\
Personal experience & 3 & \hyperlink{A4}{A4}, \hyperlink{A22}{A22}, \hyperlink{A32}{A32} \\
Survey (standalone) & 3 & \hyperlink{A27}{A27}, \hyperlink{A29}{A29}, \hyperlink{A31}{A31} \\
Survey (as follow up) & 7 & \hyperlink{A7}{A7}, \hyperlink{A9}{A9}, \hyperlink{A20}{A20}, \hyperlink{A21}{A21}, \hyperlink{A24}{A24}, \hyperlink{A30}{A30}, \hyperlink{A34}{A34}\\
Think-aloud interviews & 3 & \hyperlink{A20}{A20}, \hyperlink{A23}{A23}, \hyperlink{A30}{A30} \\
Worked example & 7 & \hyperlink{A1}{A1}, \hyperlink{A5}{A5}, \hyperlink{A6}{A6}, \hyperlink{A10}{A10}, \hyperlink{A14}{A14}, \hyperlink{A18}{A18}, \hyperlink{A32}{A32} \\
Workshop & 9 & \hyperlink{A9}{A9}, \hyperlink{A11}{A11}, \hyperlink{A20}{A20}, \hyperlink{A21}{A21}, \hyperlink{A24}{A24}, \hyperlink{A25}{A25}, \hyperlink{A27}{A27}, \hyperlink{A34}{A34}, \hyperlink{A35}{A35} \\ \bottomrule
\end{tabular}%
}
\end{table}

\subsection{Usability as the primary evaluation method}
\label{statusquo:patterns_usability}

\subsubsection{Description of current evaluation practices}
\label{statusquo:patterns_usability_description}

Choices about evaluation design, participant recruitment, and evaluation metrics are made---often implicitly---to understand the extent to which users of an RAI tool use it as the designers intended.\footnote{While outside the scope of this paper, we note that users often appropriate technologies for purposes other than their designers intend \cite{Pinch1984-lm}.} Several publications in the corpus explicitly adopt a user-centered framework [\hyperlink{A7}{A7}, \hyperlink{A15}{A15}, \hyperlink{A31}{A31}, \hyperlink{A35}{A35}] or usability focus [\hyperlink{A7}{A7}, \hyperlink{A15}{A15}, \hyperlink{A16}{A16}, \hyperlink{A20}{A20}, \hyperlink{A30}{A30}] in their evaluation of an RAI tool. In others, evaluation design choices implicitly reflect a user-centered framework: data collection methods focus on capturing how individual users interact with a tool [\hyperlink{A15}{A15}, \hyperlink{A20}{A20}, \hyperlink{A23}{A23}, \hyperlink{A28}{A28}, \hyperlink{A31}{A31}] and usability metrics are used for evaluations [\hyperlink{A21}{A21}, \hyperlink{A29}{A29}, \hyperlink{A30}{A30}]. Meanwhile, claims about tool effectiveness are only directly discussed in four publications: in [\hyperlink{A9}{A9}] and [\hyperlink{A21}{A21}] the authors study the \textit{``initial effectiveness''} of the tools they investigate through a case study of an educational workshop guide in a classroom [\hyperlink{A9}{A9}] and an empirical investigation of a workshop guide across a series of pilot workshops [\hyperlink{A21}{A21}]; and in [\hyperlink{A34}{A34}] the authors posit that the \textit{``societal impact''} of the workshop guide they describe \textit{``stems from its effectiveness as a tool to support the exploration of ethical considerations.''} Additionally, one publication in the corpus does distinguish usability from effectiveness, recognizing that \textit{``the metrics that flow directly from''} user-focused evaluation techniques \textit{``only partially capture what is instrumental to project success''} [\hyperlink{A35}{A35}]. And, several publications note in their future work or limitations sections the need for expanding evaluations to measure effectiveness \textit{``in practice''} [\hyperlink{A9}{A9}, \hyperlink{A21}{A21}, \hyperlink{A27}{A27}, \hyperlink{A31}{A31}]. Outside of these four references, no other publication in the corpus identifies effectiveness as the focus of their evaluation.\looseness=-1

\subsubsection{Gaps in existing evaluation practices}
\label{statusquo:patterns_usability_gaps}

Usability is an important consideration in the design and evaluation of RAI tools. However, RAI tools often aim to have impact in ways that may be in tension with, or outside the scope of, usability evaluations. In such cases, the results of an evaluation focused on usability may be misleading, and a focus on effectiveness may be preferable: an RAI tool may be easy to use by practitioners, but ineffective in terms of changing practitioners' behavior or the resulting product they are developing. Relatedly, as [\hyperlink{A28}{A28}] discusses, a tool's ease of use may result in over-reliance and over-trust in it. Reflecting this, across several types of RAI tools, publications describe the tool's objective as being to prompt practitioners to reflect on and disrupt their usual ways of working (e.g., via a transparency artifact [\hyperlink{A4}{A4}, \hyperlink{A25}{A25}], software toolkit [\hyperlink{A27}{A27}, \hyperlink{A30}{A30}], or workshop guide [\hyperlink{A33}{A33}, \hyperlink{A34}{A34}, \hyperlink{A36}{A36}]). In some cases, the purpose of the RAI tool is to enable consideration or inclusion of the perspectives of stakeholders or communities who may be affected by the algorithmic system being developed [\hyperlink{A4}{A4}, \hyperlink{A33}{A33}, \hyperlink{A34}{A34}]. Implicit in this purpose is the recognition that usual ways of working do not enable these perspectives to be included during AI system development. Given this, RAI tools may need to balance the needs of their users against the broader objectives of the tool in changing AI development processes towards more fair, transparent, or accountable systems. One publication [\hyperlink{A30}{A30}] recognizes this trade-off: the authors evaluate an automated tool in comparison to an interactive tool, noting that while automation \textit{``has a beneficial effect of improving efficiency''} for users, \textit{``it might come at the cost of reduced understanding''} on the part of the user \textit{``due to lack of exploration.''}\footnote{Recent HCI research has identified the value of friction and \textit{``seam-ful''} design in fostering reflection in design processes \cite{ehsan2022seamful,kaur2022sensible}.} As we discuss in Section \ref{relatedwork:evaluation_goals}, by broadening the scope of evaluation from usability to also include effectiveness, the field's evaluations of RAI tools can address both sides of this trade-off.\looseness=-1

\subsection{Fitting within work practices}
\label{statusquo:patterns_scope}

\subsubsection{Description of current evaluation practices}
\label{statusquo:patterns_scope_description}

Consistent with a focus on usability, evaluation practices focus on determining whether RAI tools are suitable for the existing work contexts of AI practitioners. This focus stems from design requirements for RAI tools and assumptions made about how practitioners approach RAI-related work. Design requirements, which, in an evaluation focused on usability, form the benchmark against which an RAI tool is evaluated, emphasize the need for RAI tools to integrate into practitioners' workflows, across a range of RAI tool types (e.g., transparency artifacts [\hyperlink{A4}{A4}], software toolkits [\hyperlink{A15}{A15}, \hyperlink{A16}{A16}, \hyperlink{A22}{A22}, \hyperlink{A27}{A27}, \hyperlink{A29}{A29}], or workshop guides [\hyperlink{A36}{A36}]). For instance, some tool publications explicitly state that their users should be able to use the tool \textit{``within their existing workflow having to write little or no code''} [\hyperlink{A16}{A16}], or modify the tool \textit{``based on their existing organizational infrastructure and workflows''} [\hyperlink{A4}{A4}]. Design requirements also assume that use of an RAI tool will occur within a fast paced work environment. RAI tools are thus framed as needing to be easy to use for busy practitioners [\hyperlink{A2}{A2}, \hyperlink{A14}{A14}, \hyperlink{A16}{A16}, \hyperlink{A19}{A19}, \hyperlink{A27}{A27}]. Tools need to enable rapid onboarding \textit{``due to workplace time constraints''} [\hyperlink{A27}{A27}]; need to be \textit{``approachable''} [\hyperlink{A14}{A14}] and \textit{``easily adopted''} [\hyperlink{A2}{A2}]; need to enable users to \textit{``try out''} features \textit{``without reading documentation''} [\hyperlink{A16}{A16}]; and need to enable users to \textit{``quickly test hypotheses and build understanding''} [\hyperlink{A19}{A19}].\looseness=-1

\subsubsection{Gaps in existing evaluation practices}
\label{statusquo:patterns_scope_gaps}

An underlying premise of RAI tools is that the default ways of developing AI systems are insufficient to lead to the goals of responsible AI (e.g., fairness, transparency, accountability) on their own. Reflecting this, researchers have focused on the ``principles to practice'' gap in responsible AI, identifying various obstacles to the translation of high-level responsible AI principles into changes in AI development practices \cite[e.g.,][]{Schiff2020-rw, Madaio2020-ou, Eitel-Porter2021-qx}. When evaluating an RAI tool, evaluators should consider whether framing the use of the RAI tool as subordinate to existing ways of working is appropriate, given the objectives of the tool under evaluation. It may be the case that an RAI tool can only be effective when existing ways of working are changed. Reflecting this, the authors of [\hyperlink{A29}{A29}] note that it is \textit{``important to consider whether toolkits with necessarily reductionist definitions of fairness are appropriate and beneficial from a societal standpoint.''} Similarly, in their discussion of future challenges, the authors of [\hyperlink{A4}{A4}] argue that, \textit{``organizational infrastructure and workflows—not to mention incentives—will need to be modified to accommodate''} meaningful adoption of an RAI tool.

Emerging work on maturity models of organizational readiness for RAI \cite[e.g.,][]{Baxter_undated-ro,hegerall,jantunen2021building,krijger2022ai,Chang2022-jc} suggests that organizational culture, incentives, and processes may matter as much as (and are likely to impact) individual practitioners' ability to use an RAI tool in their work. As such, existing ways of working should be accounted for in the evaluation of an RAI tool (and may act as a confounding factor for a given tool's effectiveness in a particular organizational context). For example, the size of a company in which an RAI tool is being used (and, similarly, the company's internal organizational structure) may have a large effect on whether or to what extent that tool is effective---e.g., smaller companies may lack the capacity to have dedicated teams or individuals focusing on RAI \cite[cf.][]{winecoff2022artificial,widder2022dislocated,Rakova2021-mc}. In fact, greater precision in investigating such questions (e.g., precisely how individual, team, or organizational factors may impact AI design processes or outcomes) is one potential result from more robust methodologies for evaluating effectiveness of RAI tools, as we explore further in Section \ref{improving:validity}.

\subsection{Replicating \textit{``real world''} settings, tasks, or users}
\label{statusquo:patterns_real_world}

\subsubsection{Description of current evaluation practices}
\label{statusquo:patterns_real_world_description}

Publications describe their evaluation designs using the terms \textit{``realistic''} [\hyperlink{A25}{A25}, \hyperlink{A26}{A26}, \hyperlink{A27}{A27}, \hyperlink{A31}{A31}] or \textit{``real world''} [\hyperlink{A1}{A1}, \hyperlink{A9}{A9}, \hyperlink{A18}{A18}, \hyperlink{A20}{A20}, \hyperlink{A27}{A27}, \hyperlink{A28}{A28}]. In particular, evaluations attempt to import three aspects of the \textit{``real world''} into the evaluation design: participants that are thought to be representative of \textit{``real world''} users [\hyperlink{A7}{A7}, \hyperlink{A20}{A20}, \hyperlink{A27}{A27}, \hyperlink{A29}{A29}] (but sometimes without an explicit definition of what representative means in this case \cite[cf.][]{chasalow2021representativeness}); activities for participants to undertake that are chosen to be reflective of the sorts of activities practitioners will use the tool for in the \textit{``real world''} [\hyperlink{A1}{A1}, \hyperlink{A4}{A4}, \hyperlink{A9}{A9}, \hyperlink{A14}{A14}, \hyperlink{A27}{A27}, \hyperlink{A28}{A28}, \hyperlink{A30}{A30}, \hyperlink{A31}{A31}]; and, \textit{``real world''} environments in which such activities are thought to occur [\hyperlink{A18}{A18}, \hyperlink{A31}{A31}]. Reflecting this, in [\hyperlink{A27}{A27}] the authors note they \textit{``designed a realistic ML task in which we required practitioners to build an ML model based on a real-world dataset.''} The use of scenarios designed to resemble \textit{``real world''} tasks is repeated across the corpus.\looseness=-1

In many cases, this means relying on publicly available datasets for the tasks. In [\hyperlink{A31}{A31}] the authors note that \textit{``we tried to put data scientists in a realistic setting,''} which consisted of a Jupyter Notebook that individual participants used to explore a \textit{``publicly available ML dataset based on 1994 US census data.''} COMPAS data and recidivism prediction scenarios are used frequently [\hyperlink{A1}{A1}, \hyperlink{A9}{A9}, \hyperlink{A14}{A14}, \hyperlink{A30}{A30}], as are the German Credit Data dataset [\hyperlink{A14}{A14}, \hyperlink{A16}{A16}, \hyperlink{A32}{A32}] and the Adult Census Income datasets [\hyperlink{A14}{A14}, \hyperlink{A22}{A22}, \hyperlink{A28}{A28}, \hyperlink{A30}{A30}, \hyperlink{A31}{A31}] in credit-worthiness and loans scenarios. Various other publicly available datasets are also used [\hyperlink{A2}{A2}, \hyperlink{A4}{A4}, \hyperlink{A6}{A6}, \hyperlink{A8}{A8}, \hyperlink{A27}{A27}, \hyperlink{A32}{A32}]. Alternatively, participants may be asked to use the RAI tool in \textit{``real-world projects''} and report back findings [\hyperlink{A18}{A18}].

Participants in evaluations are also often recruited for their verisimilitude to intended users of RAI tools. One publication notes, \textit{``We recruited and explored the perspectives of people who we believe are representative of the future data workforce''} [\hyperlink{A15}{A15}]. Several publications in the corpus describe the \textit{``real world''} users they recruit in terms of job title [\hyperlink{A21}{A21}, \hyperlink{A23}{A23}] or team membership [\hyperlink{A24}{A24}, \hyperlink{A34}{A34}], academic training [\hyperlink{A20}{A20}], or experience with certain technologies [\hyperlink{A15}{A15}, \hyperlink{A27}{A27}, \hyperlink{A31}{A31}]. Participants are often recruited through the networks of evaluators through direct outreach, snowball sampling, and outreach through online practitioner forums [\hyperlink{A21}{A21}, \hyperlink{A8}{A8}, \hyperlink{A29}{A29}, \hyperlink{A27}{A27}, \hyperlink{A28}{A28}, \hyperlink{A15}{A15}, \hyperlink{A23}{A23}, \hyperlink{A31}{A31}]. In some instances, recruitment of participants occurs exclusively in the large technology firm which the evaluation authors work at [\hyperlink{A7}{A7}, \hyperlink{A20}{A20}, \hyperlink{A31}{A31}, \hyperlink{A34}{A34}]. In other instances, recruitment efforts focus on computer science students [\hyperlink{A9}{A9}, \hyperlink{A20}{A20}, \hyperlink{A25}{A25}, \hyperlink{A30}{A30}].\looseness=-1

\subsubsection{Gaps in existing evaluation practices}
\label{statusquo:patterns_real_world_gaps}

Designers of RAI tools aspire for them to be used by practitioners in their day-to-day work. As such, evaluating tools in settings that are as close as possible to those in which practitioners work makes sense. However, evaluations of RAI tools capture only certain aspects of the contexts in which the tool is intended to be used. As such, current evaluations often fail to consider how decisions about \textit{which} aspects of the ``real world'' to capture in evaluations can impact the types of claims that can be made about how RAI tools are and will be used. Existing ``real world'' evaluations of RAI tools thus risk making claims to generalizability that are unsupported by their evaluation design. When participants in an evaluation, for instance, are recruited for their similarity to intended users of an RAI tool, evaluators must choose which aspects of intended users' identities and sociopolitical context fall within their definition of ``real world'' users. Are evaluation participants defined solely by their professional training? Doing so means that the potential for other aspects of their identity---e.g., demographics, cultural background, or organizational context---to influence their use or impact of the RAI tool cannot be established by the evaluation. One publication [\hyperlink{A9}{A9}] reports the demographic background of evaluation participants; this publication also concludes with a short discussion of additional social contexts in which the effectiveness of the RAI tool should also be evaluated. An additional study, [\hyperlink{A23}{A23}], documents its decision not to collect or report demographic information, due to concerns about framing and anchoring effects, and notes that future research should consider the relationship between user identity and tool use.

Although different methods and research communities have different standards for sample sizes \cite{caine2016local} (and this should not be used as a proxy for evaluation quality), the general reliance in evaluations of RAI tools on small sample sizes and evaluators' networks for participant recruitment, means claims that participants are representative of ``real world'' users are particularly fraught---or should at least be contextualized with greater detail about participants and their contexts, as in rigorous qualitative research \cite[e.g.,][]{Small2022-gb}. Indeed, several publications in the corpus note their small sample size as a limitation [e.g., \hyperlink{A28}{A28}, \hyperlink{A29}{A29}]]. Likewise, when tasks in an evaluation are designed to be similar to the ways intended users are expected to interact with the tool, evaluators must determine which aspects of intended use to replicate in the evaluation design, as the authors of [\hyperlink{A31}{A31}]acknowledge when discussing the limitations of their evaluation design. Will participants be allowed to use their usual computer setup, or will they need to log into an evaluation-specific environment? Will participants have a time limit to undertake the task? Will they be able to work together with teammates or other collaborators? These decisions are often described in terms of pragmatic constraints, but nonetheless need to be considered from the perspective of external validity of evaluations of RAI tools.

Indeed, the importance placed on evaluating RAI tools in the ``real world'' demonstrates the implicit value placed generally on questions of external validity in evaluations of RAI tools, and in particular, on questions of ecological validity (as one sub-type of external validity). Yet, describing an evaluation design as a ``real world'' task, without documenting the ways the design inevitably differs from actual use, risks misrepresenting the extent to which an RAI tool is ready for widespread RAI community adoption and makes determining the extent to which an evaluation has ecological validity challenging. We expand on limitations of the ``real world'' in Section \ref{improving:validity}, where we recommend reconsidering how external validity is reflected in evaluation designs.

\subsection{Individual use of RAI tools}
\label{statusquo:patterns_tool_users}

\subsubsection{Description of current evaluation practices}
\label{statusquo:patterns_tool_users_description}

Evaluations of RAI tools tend to study interactions between an individual user and the tool (e.g., between a practitioner and a software toolkit [\hyperlink{A15}{A15}, \hyperlink{A20}{A20}, \hyperlink{A28}{A28}, \hyperlink{A31}{A31}], or a practitioner and transparency artifact [\hyperlink{A23}{A23}]), despite the collaborative nature of AI development (and software development more broadly). Reflecting this, evaluations in the corpus involve activities where individual participants interact in isolation with an RAI tool, e.g. by using an interactive programming environment (e.g., Jupyter Notebook or Google Colab) to explore a software toolkit [\hyperlink{A20}{A20}, \hyperlink{A28}{A28}, \hyperlink{A31}{A31}]. This individualistic conceptualization of tool use extends to the way evaluations envision how users might learn about a tool. For the most part, evaluation designs introduce participants to the RAI tool under evaluation by inviting them to read technical documentation [\hyperlink{A29}{A29}], follow a tutorial [\hyperlink{A31}{A31}], or interact with a worked example [\hyperlink{A16}{A16}, \hyperlink{A20}{A20}, \hyperlink{A28}{A28}, \hyperlink{A31}{A31}]. A number of authors note that their evaluation designs were compromised by COVID-19 social distancing requirements [\hyperlink{A9}{A9}, \hyperlink{A28}{A28}, \hyperlink{A30}{A30}], which may have contributed to this tendency.

In contrast, a smaller subset of evaluations introduce participants to the RAI tool under evaluation through more social modes of learning, such as by participating in a workshop [\hyperlink{A7}{A7}], or by using the tool in daily work for a number of weeks [\hyperlink{A9}{A9}]. Where evaluations find that participants \textit{``misuse''} a tool, they are likely to interpret misuse as indicating flaws in the design of the tool, rather than in the onboarding or support provided to participants [\hyperlink{A27}{A27}, \hyperlink{A28}{A28}, \hyperlink{A29}{A29}]. One evaluation reports identifying \textit{``several gaps between the tools’ capabilities and the practitioners’ needs''} [\hyperlink{A29}{A29}], and another reports that \textit{``almost half of the participants were not able to provide the correct interpretation of what the model did using the }[tool]\textit{ alone''} [\hyperlink{A15}{A15}]. In some instances, such as [\hyperlink{A15}{A15}], while the evaluation activity only explores individual use of an RAI tool, the broader evaluation write-up does consider how insights from individual user-tool interactions might translate into collaborative team environments. Reflecting this, in [\hyperlink{A15}{A15}] the authors conclude their paper by observing that, \textit{``Decision-making with data and deep learning models is a collaborative and distributed process... For }[the RAI tool]\textit{ to be adopted and impact organizational processes, they must support knowledge sharing and negotiation across stakeholders}.''

\subsubsection{Gaps in existing evaluation practices}
\label{statusquo:patterns_tool_users_gaps}

As the above extract from [\hyperlink{A15}{A15}] argues, development of AI systems is a complex social activity, often undertaken in collaborative team contexts, within a web of power relations \cite[e.g.,][]{Sloane2022-wt,Petterson2023-jc,widder2023it}. RAI tools aim to intervene in this social activity, and as such are themselves likely to be used in team environments, where the intended user of an RAI tool may not be the team leader or primary decision maker. Additionally, as many social theories of learning highlight \cite[e.g.,][]{vygotsky1978mind, wenger2009social}, use of a tool is likely to occur alongside support from more experienced peers. As such, evaluation designs in which individual participants who are thought to be representative of intended users interact independently with a tool are unable to consider how the team environment may affect tool adoption or effectiveness. While some publications noted that future work should include developing training workshops or mentor networks to support tool adoption [e.g., \hyperlink{A2}{A2}], most did not consider incorporating social approaches to learning into the evaluation design itself.

An AI practitioner may find that following a protocol to produce a transparency artifact for a training dataset is easy to do during an evaluation task, but then struggle to secure their manager's support for investing time in producing such artifacts during their day-to-day work. Similarly, a practitioner may find using a software toolkit to evaluate model fairness straightforward during an evaluation task, but in practice find it difficult to take action on insights from their analysis within their broader team context \cite[cf.][]{Madaio2022-db,deng2023investigating}. Reflecting this, in one study, participants were placed in a scenario where they \textit{``were told that they would serve as the decision-maker for their team''} and were provided with a notebook containing \textit{``analysis done by one of their team members''} which the study asked them to review [\hyperlink{A28}{A28}]. Such an evaluation recognizes that the RAI tool will be used in a team setting, but operationalizes that use in a way that places the interpersonal dynamics of a team outside the scope of the evaluation: communication between a team member and the team leader is operationalized as review of a notebook; decision-making is operationalized as an individual reflective task. While this may render the evaluation more tractable, it is likely to pose threats to external validity, particularly in terms of claims about the effectiveness of the tool in team settings that rely on different communication and collaboration approaches. More generally, it may also be the case that the aim of independent RAI tool use is unrealistic. The complexity of intervening to shift AI development processes towards more responsible practices may require both individual training in particular skills, knowledge, or tools, as well as collaboration from other team members or external stakeholders.\looseness=-1

\section{Towards evaluation of RAI tool effectiveness}
\label{sec:improving_evaluation}

In Section \ref{sec:existing_practices} we highlighted patterns in existing evaluation practices. For each pattern we discussed gaps between the practice and the aims of RAI tool developers and the RAI community. To address these gaps, in this section we apply the insights from evaluation practices outside the RAI community introduced in Section \ref{sec:related_work} to propose design desiderata for the development of an \textit{effectiveness evaluation framework}, and suggest steps for RAI tool developers and the RAI and HCI community to take in support of more robust effectiveness evaluations of RAI tools.

\subsection{Design desiderata for an effectiveness evaluation framework}
\label{improving:evaluation_framework_desiderata}

The objective of an effectiveness evaluation framework is to support evaluators of RAI tools to design robust evaluations, commensurate to similar efforts in the HCI and software engineering fields to develop frameworks for usability evaluations \cite[e.g.,][]{ko2015practical, davis2023what}, evaluations of ML models \cite{Hutchinson2022-gh}, and the design of empirical studies in software engineering \cite[e.g.,][]{kitchenham2002preliminary}. Below we provide initial design desiderata for the development of a framework that meets this objective.\looseness=-1

It is important to note that RAI tools are often evaluated within the context of an ongoing, resource-constrained tool development process. 
As such, an effectiveness evaluation framework cannot exist in isolation from the processes by which RAI tools are developed. For instance, the resource constraints facing those developing RAI tools may need to be considered---as it is likely those evaluating RAI tools will face similar constraints (indeed, they may often be the same teams). Therefore, an effectiveness evaluation framework needs to enable evaluators to reflect on potential trade-offs or limitations they face in conducting evaluations of RAI tools, and to reflect on their own relationship to the developers and other stakeholders of the RAI tool under evaluation. Additionally, as discussed in Section \ref{methods:identifying_evals}, the intended users of RAI tools are often AI practitioners. Accessing practitioners and observing their work environments is a well-documented challenge \cite[e.g.,][]{wang2023designing, varanasi2023it}, as reflected in the tendency of existing evaluation practices to use mock scenarios and recruit participants from a single workplace (Section \ref{statusquo:patterns_real_world}). An effectiveness evaluation framework, therefore, must provide evaluators with guidance even in situations where participant recruitment or engagement may be challenging.

An effectiveness evaluation framework must also offer clear guidance on issues of internal and external validity, and trade-offs between these, as the expanded scope of questions of effectiveness, compared to questions of usability, heightens the importance of careful justification of evaluation claims. A claim that a tool is easy to use has far narrower scope, and less significant consequences if invalid, than a claim that a tool is effective at addressing a particular ethical issue during AI development. In this regard, the discussion of external validity and best practices in education evaluations in Section \ref{relatedwork:evaluation_validity} can provide guidance for the development of an effectiveness evaluation framework. In particular, the emphasis placed in other fields on documentation requirements for evaluation is instructive (see the discussion in Section \ref{relatedwork:evaluation_goals} on the IES's What Works Clearinghouse). Detailed documentation of evaluation design, participant recruitment and demographics, and evaluation limitations is needed, as all of these have implications for the broader set of use cases, contexts, and populations for which evaluation findings are likely to be relevant, as with external validity of evaluations in other fields \cite[e.g.,][]{towne2002scientific,gertler2016impact}. Ideally, evaluations should be documented with enough detail to enable RAI community members who are considering adopting an RAI tool to determine whether evaluation settings are likely to be comparable to their own context.

Further work is required to develop these desiderata into a practical framework to guide evaluations of RAI tools. This work will need to be collaborative, engaging RAI tool designers, users, and stakeholders, and drawing on the expertise of the RAI community. Any framework will need to be maintained and updated over time. The What Works Clearinghouse's Procedures and Standards Handbook \cite{whatworks2022} provides one example of a highly formalised approach to framework development and maintenance. Northwestern University's Machine Learning Impact Initiative, which produced a framework for evaluation of ML applications \cite{hammond2021framework}, provides another example of framework development within a collaborative academic context. Finally, the development of a sociotechnical framework for explainable AI by \citet{ehsan2023charting} provides an additional example of framework development and evaluation through a long-term industry-academic partnership. In the intervening period, however, RAI tool evaluations can be improved, in terms of their robustness and focus on effectiveness, by attending to the validity concerns outlined in Section \ref{relatedwork:evaluation_validity}.

\subsection{Addressing validity issues in RAI tool evaluations}
\label{improving:validity}

This section builds on the validity concerns described in Section \ref{relatedwork:evaluation_validity}, and offers practical guidance on addressing threats to validity during RAI tool evaluation. We note that issues of internal and external validity are intertwined, and addressing one can at times exacerbate the other. As such, we outline steps for improving internal and external validity, and then discuss how trade-offs between them can be managed during evaluation design. Across this section, we use the example of a dataset transparency protocol to illustrate our suggestions. In this context, we conceive of a dataset transparency protocol as an RAI tool designed to enable data scientists to document relevant aspects of a dataset in a standardized manner \cite[e.g.,][]{Bender2018-he, Gebru2021-ot, Pushkarna2022-sp}.\looseness=-1

\subsubsection{Improving internal validity within RAI tool evaluations}

\textit{Internal validity} refers to the aspects of evaluation design that support claims of a causal relationship between variables (i.e., between use of an RAI tool and some desired change in AI development processes). As such, two prerequisites for internal validity in an evaluation are: (1) clarity on what the evaluation is attempting to measure, and (2) clarity on any potential confounding factors. Reflecting this, software engineering and HCI guides for experimental or evaluation design begin by recommending that evaluators clearly define the evaluation or research question which their design will attempt to answer \cite[e.g.,][]{ko2015practical, kitchenham2002preliminary, greenberg2008usability, Ledo2018-yt}. In the context of evaluations of RAI tools, internal validity is difficult to establish when there is a lack of clarity on the goals of the tool itself, or on how effectiveness of the tool is defined. A dataset transparency protocol, for example, could have a goal of improving dataset curation practices, or of reducing misuse of datasets in ML training processes. For each of these goals, effectiveness can be defined in multiple ways, and different confounding factors may be relevant. An additional challenge to internal validity is the dynamic and highly interdependent nature of AI development processes, which can confound attempts at establishing a causal relationship between RAI tool use and desired outcomes. Given this, for the purposes of improving internal validity, what matters is that evaluators are specific about the particular goals against which they are evaluating an RAI tool.\looseness=-1

RAI tool evaluators can begin to address internal validity in two ways. First, evaluators can increase documentation of the confounding variables that may affect any claims they make as to tool effectiveness. Whilst this will not mitigate the impact of confounding variables, it will enable readers of an evaluation to better interpret evaluation results. Second, RAI tool evaluators can articulate a \textit{theory of change} for the effectiveness of the RAI tool, which can be used to guide identification of confounding variables and prioritise their mitigation. A theory of change is a working hypothesis for how adoption and (intended) use of a given tool will achieve the tool's goals.\footnote{See \citet{Wong2022-zi} and the FAccT 2023 CRAFT session on theories of change in Responsible AI for more on this: \url{https://www.youtube.com/watch?v=DyDLnJIwa08}.} The theory of change explains, in a sequence of causal statements, precisely how use of the tool in a specific context may achieve the desired goals. Ideally, each causal statement in a theory of change will be evaluated. Where this is not possible, identifying and documenting the assumptions underpinning a given tool will enhance the ability of third-parties to determine for themselves whether the tool is likely to be applicable in their context.\looseness=-1

As an illustrative example, for a dataset transparency protocol, one goal may be to ensure downstream users of the dataset know how to use it for tasks or contexts for which the dataset is appropriate. A corresponding theory of change may be: \textit{If} documentation creators use the tool to record relevant characteristics of datasets, and \textit{if} practitioners review a dataset's documentation before deciding to use it, \textit{then} training datasets will only be used in training tasks for which they are appropriate. Pertinent confounding variables this theory of change reveals include: documentation creators' expertise or lived experience, which may affect how they use the tool and what characteristics of the dataset they identify as relevant\footnote{One approach to identifying such confounding variables may be to draw on sense-making theory, as \citet{kaur2022sensible} explore in their discussion of how the identity and social context of the person receiving a model output shapes their interpretation of the output.}; and, the various factors practitioners must balance when determining whether to use a given dataset (e.g., availability of alternative datasets).

\subsubsection{Enhancing the external validity of RAI tool evaluations}

\textit{External validity}, in the context of an RAI tool evaluation, refers to the extent to which evaluation findings may be generalized to new contexts. The extent to which external validity is a concern for RAI tool evaluators largely depends on the sorts of claims they wish to make about the tool. In a usability study, external validity can be enhanced through careful documentation of study participants, their demographics, their existing skills, and the training or support they received during evaluation activities---each of which may constrain external validity. In an evaluation focused on determining RAI tool effectiveness, however, these steps are insufficient as the scope of the evaluation is inherently broader. Here, conceptualizing RAI tools as interventions in AI system development and deployment, as we discuss in Section \ref{relatedwork:RAI_interventions}, is instructive: the effectiveness of an intervention is inherently tied to the social context in which it occurs. As such, to address external validity in an effectiveness evaluation requires RAI tool evaluators to consider the role of social context in shaping the outcomes of RAI tool use. In other words, external validity requires RAI tool evaluators to ask: to what extent is the apparent success of this RAI tool a result of social factors (e.g., pressure from management, high buy-in from evaluation participants, a workplace culture that emphasises ethics), rather than tool characteristics?\looseness=-1

As with internal validity, the theory of change behind an RAI tool can serve as a useful starting point for attending to external validity in RAI tool evaluation. Returning to our illustrative example, for a dataset transparency protocol, we outlined two causal links in the theory of change (there are likely others too): documentation creators must use the tool correctly, and practitioners must review dataset documentation when making decisions about using the dataset. Both of these links present threats to external validity. Snowball or convenience recruitment methods, for example, might result in a sample of participants who are far more highly engaged in RAI issues and discourse than the target user group for the tool, leading to evaluators over-estimating the effectiveness of their tool.\footnote{See \citet{baltes2022sampling} for a longer discussion of mitigation strategies for validity threats introduced by widely used sampling strategies in qualitative software engineering research.} To mitigate this, the social context associated with each causal statement in the theory of change can be documented as a limitation to external validity, or can be used to expand the scope of evaluation practices---i.e., from largely observational studies with individual practitioners using the tool in in-lab studies or single-site case studies (see Table \ref{tab:eval_activities}). For instance, evaluations of effectiveness might instead adopt wider units of analysis: e.g., from individual users to team-level analyses; from single site to multi-site comparative studies.\footnote{Moving beyond individual users may also be important for usability evaluations, particularly where the tool is designed to be used collaboratively \cite{Grudin1988-pt}.} 

New evaluation approaches may be needed to support these team-level analyses as well as cross-context comparative studies. Approaches from the social sciences, such as educational intervention evaluations \cite[e.g.,][]{towne2002scientific,clearinghouse2012works} and public health or economic policy evaluations \cite[e.g.,][]{gertler2016impact} may be constructively adopted by the RAI community. At a community level, such approaches may offer evidence of the positive impact of robust evaluation practices; at an individual evaluator level, such approaches may offer guidance on implementation of new evaluation methods. Additionally, recent work has explored how existing algorithmic fairness paradigms originating in Western contexts may not be applicable in the global contexts of AI systems' use \cite[e.g.,][]{goyal2022your, sambasivan2021re,ghosh2021detecting,bhatt2022re,prabhakaran2022cultural}. Such work is likely to be fruitful in informing more comparative, cross-context studies evaluating the effectiveness of RAI tools in different cultural contexts, for instance, by indicating aspects of RAI tool use that are likely to change from context to context, and highlighting culturally-specific assumptions encoded in tool designs.

An important aspect of external validity is \textit{ecological validity}, which is the extent to which an RAI tool evaluation is reflective of a ``real world''. Indeed, addressing ecological validity may prompt RAI tool evaluators to reconsider how evaluation settings relate to the ``real world''. When RAI researchers discuss ``real world'' studies, it should be clear which aspects of the world (or even which \textit{world} \cite[cf.,][]{escobar2018designs,Unruh1980-gg}) their study is intended to emulate. The distinction between implicitly synthetic or artificial laboratory settings and the messy reality of the everyday may be helpful, insofar as it invites reflection on differences between the context in which an RAI tool is developed and tested \cite{ribes2019logic}, and the contexts in which the tool will be deployed \cite{GOYAL2016EFFECTS}. However, the contrast risks making the same positivist assumption as that underpinning the distinction between so-called ``raw'' and ``processed'' data: that there is a single, shared world (i.e., a world in which data exists in its raw state, separate from its situatedness in sociocultural systems). This assumption is ill-suited to the task of evaluating RAI tools for two reasons: first, in its focus on a singular world, the assumption downplays the role of culture in co-constructing social worlds, including in co-constructing the ways users interpret and interact with RAI tools and AI systems. Second, the conflation of the ``real world'' with the world of intended RAI tool users, who are overwhelmingly AI practitioners working in WEIRD (Western, educated, industrialized, rich, democratic) contexts \cite[see, for a discussion of this acronym,][]{Henrich2010-tl}, obfuscates critical discussion about the ways that WEIRD norms and logics may shape the adoption of RAI tools.\footnote{See \citet{Van_Berkel2023-xh} and \citet{septiandri2023weird} for an analysis of the WEIRDness of participants in HCI and FAccT, and \cite{prabhakaran2022cultural,dev2022measures, sambasivan2021re,ghosh2021detecting,bhatt2022re} for examples of cross-cultural evaluations in RAI.}

Additionally, when participants in RAI evaluations are described as ``real world'' practitioners (see Section \ref{statusquo:patterns_real_world_description}), the downplaying of cultural context and tacit centering of WEIRD norms can be seen in the aspects of participants---such as, their cultural background and physical location---that are (not) documented. Here, Haraway's \textit{god trick} \cite{Haraway1988-lt} is in full effect: evaluation participants representing only a narrow slice of \textit{a} world are rendered as universal (and therefore their interactions with an RAI tool as objective and generalizable) by representing them as unsituated and placeless. Indeed, beyond greater reflection on the choices implicit in operationalizing the ``real world'' in the design of evaluations, development and adoption of an effectiveness evaluation framework may result in evaluators abandoning this fiction in evaluation designs altogether. Instead, the external validity of evaluations may be improved by adopting more modest aims, such as evaluating the effectiveness of an RAI tool in a specific practice and context.

\subsubsection{Addressing trade-offs between internal and external validity}
\label{improving:validity_tradeoffs}

As discussed in Section \ref{improving:evaluation_framework_desiderata}, building on existing usability evaluation practices to also consider evaluation of RAI tool effectiveness will necessarily amplify threats to internal and external validity. For evaluators, responding to this is complicated by the fact that prioritizing one type of validity often leads to concessions on the other type. In the example of a dataset transparency protocol, for example, prioritizing internal validity may lead to a highly controlled in-lab evaluation task (e.g., where participants interact with an example dataset and scenario, following structured prompts). Here, the impact of variations in participants' expertise or lived experience can be managed through random assignment of participants to control and treatment groups (i.e., a group who does not use the dataset transparency protocol and a group that does). Such an evaluation design, however, will necessarily have limited generalizability, due to the artificiality of the evaluation set up. Conversely, prioritizing external validity in the design of an evaluation of a dataset transparency protocol may lead to a longitudinal study of day-to-day protocol use in multiple sites, across multiple teams. This design will necessarily have weak internal validity due to the number of confounding variables, but may have strong external validity if participant recruitment and site selection are carefully managed. 

In general, narrowing the breadth of evaluation claims, particularly regarding generalizability, will also narrow the breadth of external validity issues, thereby reducing pressure on the trade-offs between internal and external validity. In the dataset transparency protocol example, we might adopt a snowball sampling strategy, but then include in our evaluation write-up a summary of participants' professional training or experience (e.g., as in \cite{Crisan2022-zz, Boyd2021-dr}), and a discussion of the extent to which participants' backgrounds is likely to be representative of the backgrounds of intended tool users. However, given the breadth of different contexts in which RAI tools are used, the ongoing evolving nature of AI development practices, and resource constraints facing evaluators, it may be impossible to meaningfully address all confounding variables in an evaluation design. In such a situation, one productive approach may be to refocus evaluation efforts away from determining the overarching effectiveness of an RAI tool and towards identifying \textit{indicators} for effective use of such a tool. Here, confounding factors, such as the overarching culture or risk approach in a workplace where a tool is trialed, are reconceptualized as potential indicators of effectiveness, enabling evaluation activities to focus on surfacing all confounding factors (e.g., through participant observation of in-situ tool use), rather than on attempting to control for such factors. Future work will be required to rigorously identify such indicators and develop appropriate measures. Efforts in software engineering to identify capability measures that are predictive of effective software development processes may be a useful resource to support this work \cite[e.g.,][]{monteiro2011defining, elemam2000validating}, although, as these efforts demonstrate, a field-level approach will be required to ensure any capability measures are valid across the wide range of organizational and culture settings in which RAI tools are intended to be used.

\subsection{Field-level initiatives to support RAI tool evaluations}
\label{improving:effectiveness_framework_field_changes}

RAI tool developers and evaluators face significant resource constraints, and organizational settings that may not incentivize or reward RAI work \cite{Rakova2021-mc, varanasi2023it} or may actively disincentivize it \cite{Madaio2020-ou}. While working towards an effectiveness evaluation framework that responds to these constraints, we must also work towards improving the resources devoted to RAI tool evaluation. In this Section we identify three field-level challenges to more robust effectiveness evaluations of RAI tools, and propose field-level responses to these.

\subsubsection{Identifying appropriate metrics for effectiveness evaluations.}

Operationalizing effectiveness through the design of evaluations targeted at each causal link in a theory of change will require the development of new measures for effectiveness. For instance, if researchers evaluating an RAI tool articulate a theory of change that involves the tool changing development practices in particular ways, which are then intended to change the system outcomes in particular ways, each of those steps might have their own measures that can be tracked as part of an evaluation---as well as their own confounding variables. In an experimental study (e.g., such as a randomized controlled trial), one might investigate the causal effect of introducing an RAI tool by comparing such measures from teams who used those tools with a relevant control group. However, the RAI community does not yet have a clearly defined set of what such process- or outcome-level measures might be (nor how to think about what might constitute a control group).\footnote{Some work has explored measures of specific aspects of fairness, such as disaggregated evaluations of models' performance disparities \cite{barocas2021designing, Madaio2022-db} and measuring representational harms in image captioning \cite{wang2022measuring} and natural language processing \cite{jacobs2020meaning,dev2022measures}, although this leaves out other aspects of RAI, such as transparency, accountability, privacy, and broader questions of equity and justice \cite{hoffmann2019fairness}. Moreover, these measures are largely outcome-based measures of fairness, and may not capture other notions of procedural justice \cite{lee2019procedural} or impacts on development processes.}

An additional challenge is developing process- and outcome-level measures that can be consistently operationalised across RAI tools and their intended user groups. Critics of the role that `learning outcomes' play as the metric \textit{de rigueur} in evaluating educational interventions, however, highlight three risks arising from an overly narrow focus on a single measure \cite{hussey2002trouble}: learning outcomes can appear specific, but in reducing complex phenomena into discrete outcomes can become detached from the phenomena (i.e., learning outcomes might measure test-taking skills, not learning); learning outcomes can be insensitive to the differences between disciplines; and, enforcing learning outcomes as the only evaluation metric can have the unintended consequence of suppressing innovation \cite[cf.][]{cardona2023artificial}. To mitigate these risks, \citet{hussey2003uses} have developed the concepts of `predicted/unpredicted' and `desirable/undesirable' learning outcomes, arguing that this enables a more flexible approach to evaluation. Analogously, for each stage of the AI pipeline, existing research \cite[e.g.,][]{black2023operationalizing, Rakova2021-mc} may be able to identify a set of desirable or undesirable RAI practices. Field-level alignment on appropriate proxies for these practices could be a useful first step towards developing a consistent approach to RAI tool evaluation, with RAI tool developers identifying which RAI practices they predict will be improved, and RAI tool evaluators determining the predicted and un-predicted changes in RAI practices as a result of tool adoption.\looseness=-1

Alternatively, evaluation designers might look to prior research on ``process improvement'' in software development \cite[e.g.,][]{unterkalmsteiner2011evaluation} for potential indicators of changes in development processes---however, some indicators from that line of work, such as product quality, might need to be adapted for responsible AI, such as by developing definitions of `quality' that reflect impacts on marginalized communities and society \cite[cf.][]{Raji2022-by}. Similarly, researchers might look to prior research on designing user experiments for software tools \cite[e.g.,][]{ko2015practical} for inspiration for such process-level indicators of effectiveness. However, the question of which \textit{outcome} measures may be most relevant for RAI tool evaluations (e.g., aspects of the AI system that might be changed by the use of an RAI tool; or the resulting fairness-related harms to impacted stakeholders) will require serious reflection and substantial efforts from the fields of HCI and RAI.\looseness=-1

\subsubsection{Working with stakeholders to determine 'effectiveness'.}

In widening the scope of RAI tool evaluations we should consider the ways evaluation practices signal who and what is valued in the RAI field \cite[cf.][]{Lamont2012-gl}. Who decides when an external RAI audit and review framework has led to improvements in an AI development process, and, by extension, to improvements to the AI systems developed? The goals of an RAI tool may help evaluators identify the relevant stakeholders in its evaluation. For an RAI tool whose goals are related to mitigating the impact of discriminatory AI systems, for instance, evaluation may require involving stakeholders beyond AI practitioners (e.g., members of communities impacted by discriminatory AI systems) in determining whether and in what ways the tool is effective, such as by contributing to designing and conducting the evaluation.

Some RAI tools are targeted at broader audiences than just AI practitioners and (perhaps as a result of this broader set of intended users) are also evaluated with a wider set of stakeholders \cite[e.g.,][]{Data_Cards_Playbook-ws}. This, as well as recent work calling for more, and deeper, participatory approaches to the design of AI systems \cite[e.g.,][]{icml2020wkshp,delgado2023participatory, abouk2021immediate, Birhane2022-gc, Sloane2022-sn} and involving everyday users in ``crowd audits'' of AI systems \cite[e.g.,][]{shen2021everyday, devos2022toward, deng2022understanding, lam2022end} suggest opportunities for widening the lens of who may be involved in evaluating the effectiveness of RAI tools. However, open questions remain about precisely how to involve such stakeholders in designing or conducting such evaluations---although there are lessons to be learned from participatory action research and community-based participatory research \cite[e.g.,][]{unertl2016integrating,wallerstein2006using,Hayes2014-hz,rasmussen2004action,elliot1991action,harrington2019deconstructing,Cooper2022-vz}. For instance, how might impacted community members contribute to determining what it means for an RAI tool to effectively lead to more responsible AI systems for \textit{their} community?\looseness=-1

\subsubsection{Developing evaluation frameworks, norms and standards.}

Translating the desiderata in Section \ref{improving:evaluation_framework_desiderata} into an effectiveness evaluation framework, along with evaluation guidelines and methods for researchers to apply in their evaluation study designs, will require support from the RAI---and broader HCI---community. In part, support might help expand the evaluation toolbox available to researchers developing, evaluating, or otherwise studying RAI toolkits \cite[e.g.,][]{Wong2022-zi,Petterson2023-jc,weerts2023fairlearn,kaur2022sensible}, allowing them to target their evaluations for particular goals and choose the methods appropriate to those evaluation goals. For instance, descriptive, observational studies that explore how such tools or interventions are changing practices, can be complemented by experimental or quasi-experimental studies that identify causal effects of an intervention \cite[cf.][]{gertler2016impact}.

More generally, since any individual researcher may not have the expertise, capacity, or incentives to conduct longitudinal, multi-site evaluation studies, the RAI community might provide opportunities to support or incentivize such work.\footnote{\citet{Bender2018-he} also consider similar field-level incentives, in the context of RAI tool adoption.} This might include workshops, tutorials, or special tracks at conferences to share best practices for intervention evaluation methods. In HCI, for instance, the ``RepliCHI'' special interest group was convened in response to the replicability crisis in psychology and related fields \cite{ioannidis2005most}, as a way to foster replication studies (and to focus on replicability as a desirable aspect of HCI research) \cite{wilson2012replichi, wilson2013replichi}, while the 2023 CHI conference saw the first Special Interest Group on ``Human-Centered Responsible AI'' \cite{tahaei2023human}. Additionally, HCI researchers have proposed addressing participant recruitment and design replicability challenges through the development of shared evaluation platforms, which might include reusable evaluation design guides and participant recruitment pools \cite{davis2023what}.

In addition, incentivizing such evaluations might require addressing the profit motive underlying technology firms' investments in responsible AI work \cite{Young2022-pi} through support of new centers or clusters of research institutions, multi-stakeholder initiatives (e.g., Partnership on AI), or the involvement of government agencies (e.g., the U.S.\ National Institute of Standards and Technology) or inter-governmental organizations. For instance, the World Bank has created the Development Impact Evaluation (DIME) \cite{DIME-ar} and the Strategic Impact Evaluation Fund (SIEF) \cite{SIEF-an} to fund and incentivize robust impact evaluations of development initiatives. As previously mentioned, the U.S.\ Department of Education's Institute for Education Sciences has developed the What Works Clearinghouse \cite{clearinghouse2012works} to aggregate evaluations of particular educational interventions (e.g., curricula, programs, educational models) and provide consistent standards for reporting and evaluating the robustness (including internal and external validity) of such evaluations. While not a clearinghouse, the International Organization for Standardization (ISO) and International Electrotechnical Commission (IEC) have similarly developed standards for software process evaluation \cite{elemam2000validating}. An additional model the RAI community might look to is the Open Science Framework \cite{OSF-gr}, which provides a consistent framework and publicly accessible platform for reporting details of research studies, and which is widely used to ``pre-register'' hypotheses before conducting an experiment \cite{nosek2018preregistration}, to avoid potential threats to validity from hypothesizing after results are known \cite{kerr1998harking}. What might such efforts look like for the evaluation of RAI tools?\looseness=-1

\section{Limitations}
\label{sec:limitations}

In this paper, we conduct an inductive thematic analysis of a purposive sample of publications discussing RAI tools and their evaluation. Several limitations are inherent in the purposive sampling strategy we adopted, which impact our ability to make representative claims about RAI tools and evaluation practices. Our sampling strategy did not consider regional or cultural background of tool creators, as this information was rarely included in publications, and similarly our search for RAI tools was limited to English-language publications. As such, our study cannot shed light on potential differences in evaluation practices across regions or cultures. We recognize that this is a significant limitation, as existing work highlights the importance of regional contexts in the formal and informal evaluation and regulation of RAI practices \cite{sambasivan2021re}. Our sampling strategy also focused on breadth of RAI tools, rather than on breadth of evaluation methods. Evaluation activities contained within our corpus were largely qualitative (as shown in Table \ref{tab:eval_activities}), and as such qualitative methods are the primary focus of our findings and discussion. Additionally, although publications are a significant component of HCI and ML research, and a primary way of sharing research within academia, they do not reflect the entirety of RAI tool evaluation practices (e.g., if they are conducted and shared internally, but not published). It is possible that evaluations of the RAI tools we analyze have taken place, but they have not been publicly reported. Finally, our review of evaluation methods from the HCI and education fields is not exhaustive. We draw from these fields to highlight the range of evaluation methods and resources which could be adapted for RAI tool evaluations. We recognise, however, that evaluation approaches in these fields are also debated and continue to evolve \cite[cf., regarding randomized controlled trials for medical interventions][]{cartwright2010limitations}.

\section{Conclusion}
\label{sec:conclusion}

Responsible AI tools are envisioned to lead towards more responsible AI development practices and systems. We conducted a qualitative analysis of publications that discuss RAI tools to understand current practices for their evaluation. We find that the field has focused on evaluating the usability of RAI tools, rather than on evaluating the effectiveness of those tools in changing development practices and outcomes. We identify gaps between the scope of existing evaluation practices and the ambition of RAI tools, particularly in terms of evaluation design, tasks, and participant selection. To address these, we propose the development of an \textit{effectiveness evaluation framework} for the RAI field, informed by best practices from intervention evaluations in education and medicine. We consider design desiderata for such a framework, and offer initial guidance for RAI tool evaluators to improve evaluation validity. Finally, we offer ideas for how the HCI and RAI communities might support more robust evaluations of the effectiveness of RAI tools in intervening in AI system development to lead towards more responsible AI.\looseness=-1

\begin{acks}
Thank you to Ben Hutchinson, who was an early supporter of this research project and provided invaluable advice, and to colleagues at Google Research, and the anonymous reviewers for their feedback and deep engagement with this work.
\end{acks}


\bibliography{references}
\bibliographystyle{ACM-Reference-Format}


\appendix

\section{List of RAI tools, with their primary publication}
\label{app:toollist}

\label{tab:toollist}
\begin{enumerate}
\item[T1]\hypertarget{T1}{algofairness}: benchmarking of fairness aware algorithms, \textit{discussed in publication} \hyperlink{A1}{A1}
\item[T2]\hypertarget{T2}{Data Statements}: documentation standard for reporting NLP models, \hyperlink{A2}{A2}
\item[T3]\hypertarget{T3}{BOLD}: dataset of prompts for language models to test biases, \hyperlink{A3}{A3}
\item[T4]\hypertarget{T4}{Data Sheets}: proposal for standard transparency documentation to support sharing of datasets, \hyperlink{A4}{A4}, \hyperlink{A23}{A23}
\item[T5]\hypertarget{T5}{Dataset Deprecation Report}: documentation tool for recording deprecation of a dataset from use in ML training, \hyperlink{A5}{A5}\looseness=-1
\item[T6]\hypertarget{T6}{Model Cards}: documentation standard for reporting genealogy and performance of ML model, \hyperlink{A6}{A6}, \hyperlink{A25}{A25}
\item[T7]\hypertarget{7}{Data Cards}: transparency document for data sets used in ML, \hyperlink{A7}{A7}
\item[T8]\hypertarget{T8}{Healthsheet}: application of Data Sheets for health domains, \hyperlink{A8}{A8}
\item[T9]\hypertarget{T9}{Value Cards}: toolkit for educators to engage students in challenges of designing fair ML systrems, \hyperlink{A9}{A9}
\item[T10]\hypertarget{T10}{Responsible AI License}: end user agreement for developers to restrict how AI systems can be used by others, \hyperlink{A10}{A10}\looseness=-1
\item[T11]\hypertarget{T11}{Algorithmic Equity Toolkit}: tool for community groups to challenge use of AI by government agencies, \hyperlink{A11}{A11}, \hyperlink{A35}{A35}\looseness=-1
\item[T12]\hypertarget{T12}{Algorithmic Impact Assessments}: process for public agencies to review systems they are procuring, \hyperlink{A12}{A12}
\item[T13]\hypertarget{T13}{Fairlearn}: software tool for mitigating fairness issues through a binary classification setting, \hyperlink{A13}{A13}, \hyperlink{A26}{A26}, \hyperlink{A27}{A27}, \hyperlink{A29}{A29}
\item[T14]\hypertarget{T14}{AI Fairness 360}: software toolkit for evaluation of models and datasets, \hyperlink{A14}{A14}, \hyperlink{A27}{A27}, \hyperlink{A29}{A29}, \hyperlink{A30}{A30}
\item[T15]\hypertarget{T15}{Interactive Model Cards}: software tool for interactively developing and exploring a Model Card, \hyperlink{A15}{A15}
\item[T16]\hypertarget{T16}{SageMaker Clarify}: software tool for detecting and explaining bias in models, \hyperlink{A16}{A16}
\item[T17]\hypertarget{T17}{InterpretML}: software for interpreting/explaining model outputs. Microsoft product, now open source, \hyperlink{A17}{A17}, \hyperlink{A31}{A31}\looseness=-1
\item[T18]\hypertarget{T18}{Aequitas}: software for testing models against several bias and fairness metrics, \hyperlink{A18}{A18}, \hyperlink{A28}{A28}, \hyperlink{A29}{A29}
\item[T19]\hypertarget{T19}{Language Interpretability Tool}: open source software for visualising how NLP models function, \hyperlink{A19}{A19}\looseness=-1
\item[T20]\hypertarget{T20}{What-If Tool}: software for visualizing counterfactuals of ML models without writing code, \hyperlink{A20}{A20}, \hyperlink{A29}{A29}, \hyperlink{A30}{A30}\looseness=-1
\item[T21]\hypertarget{T21}{Model Card Authoring Toolkit}: workshop guide for participatory evaluation of ML models, \hyperlink{A21}{A21}
\item[T22]\hypertarget{T22}{LiFT}: software tool for measuring biases in datasets and models, \hyperlink{A22}{A22}
\item[T23]\hypertarget{T23}{Algorithmic Accountability Policy Toolkit}: guide to ML accountability issues for legal and policy advocates, \hyperlink{A37}{A37}
\item[T24]\hypertarget{T24}{audit-AI}: software tool for exploring biases in datasets and ML models, \hyperlink{A32}{A32}
\item[T25]\hypertarget{T25}{Ethical Compass}: tool for designers to identify ethical issues, \hyperlink{A33}{A33}
\item[T26]\hypertarget{T26}{Judgment Call}: scenario game to help designers make better decisions, \hyperlink{A34}{A34}
\item[T27]\hypertarget{T27}{Consequence Scanning}: workshop for reflection on social consequences of product deployment, \hyperlink{A24}{A24}, \hyperlink{A36}{A36}\looseness=-1
\end{enumerate}

\section{RAI tools listed by target stage of AI development}
\label{app:toolcats}

The description below explains how RAI tools in the corpus are categorised into the stages of AI development, as shown in Table \ref{tab:stages}.

\begin{itemize}
    \item Two tools are designed to intervene in the problem formulation stage of AI development: \hyperlink{T12}{T12}, a tool for public agencies to review AI systems during procurement; and, \hyperlink{T27}{T27}, a facilitation guide for AI development teams to consider the consequences of product development.
    \item Two tools are designed to intervene in the design phase: \hyperlink{T25}{T25}, a tool for development teams to identify ethical issues during system design; and, \hyperlink{T26}{T26}, a guide to running workshops to help system designers make better decisions throughout system design and development. 
    \item Four tools are designed to intervene in the data collection and processing stage: \hyperlink{T4}{T4}, \hyperlink{T7}{T7}, and \hyperlink{T8}{T8}, which are protocols for creating dataset transparency artifacts; and, \hyperlink{T5}{T5}, a template for documenting decisions to withdraw a dataset from use in model training. 
    \item Two tools support interventions in the training stage: \hyperlink{T13}{T13}, a software toolkit for mitigating fairness issues in binary classification models; and \hyperlink{T14}{T14}, a software toolkit for mitigating bias in training datasets and models. 
    \item Nine tools are designed to be used during the testing stage (which often happens in iterative loops with the training stage): \hyperlink{T1}{T1}, a benchmark for comparing fairness-aware algorithms; \hyperlink{T3}{T3}, a dataset of prompts to use in bias testing of natural language models; \hyperlink{T16}{T16}, \hyperlink{T18}{T18} and \hyperlink{T22}{T22}, software toolkits for testing models against several bias and fairness metrics; \hyperlink{T17}{T17}, a software toolkit for interpreting and explaining model outputs; \hyperlink{T19}{T19}, a software tool for visualizing and interpreting  natural language models; \hyperlink{T20}{T20}, a software tool for visually probing model behavior; and, \hyperlink{T21}{T21}, a workshop guide for participatory evaluation of ML models. 
    \item Four tools are focused on the deployment stage: \hyperlink{T10}{T10}, a template for end-user agreements to restrict how AI systems can be used by third-parties; \hyperlink{T2}{T2} and \hyperlink{T6}{T6}, which are protocols for documenting the characteristics of ML models; and, \hyperlink{T15}{T15}, a software tool for interactive development of a Model Card transparency artifact. 
    \item Three tools support the monitoring stage of AI development, which we define broadly to include both first-party monitoring of deployed systems by their developers, as well as third-party reviews or audits of deployed AI systems: \hyperlink{T23}{T23}, a toolkit for policy advocates to uncover the use of AI systems and develop accountability mechanism; \hyperlink{T24}{T24}, a software toolkit focused on auditing models; and, \hyperlink{T11}{T11}, a toolkit for community groups to challenge the use of AI systems by government agencies (note, this toolkit may be used throughout the design and development pipeline, but is intended primarily for community groups to review deployed systems).
\end{itemize}
Finally, one tool in the corpus does not fit any of the aforementioned AI development stages: \hyperlink{T9}{T9}, which is an educational tool to engage students learning about the challenges of designing `fair' AI systems. 

\section{List of publications, with their associated RAI tools}
\label{app:publist}

\label{tab:publist}
\begin{enumerate}
\item[A1]\hypertarget{A1}{algofairness} (\hyperlink{T1}{T1}): \citet{Friedler2019-sc}
\item[A2]\hypertarget{A2}{Data Statements} (\hyperlink{T2}{T2}): \citet{Bender2018-he}
\item[A3]\hypertarget{A3}{BOLD} (\hyperlink{T3}{T3}): \citet{Dhamala2021-xe}
\item[A4]\hypertarget{A4}{Data Sheets} (\hyperlink{T4}{T4}): \citet{Gebru2021-ot}
\item[A5]\hypertarget{A5}{Dataset Deprecation Report} (\hyperlink{T5}{T5}): \citet{Luccioni2022-ow}
\item[A6]\hypertarget{A6}{Model Cards} (\hyperlink{T6}{T6}): \citet{Mitchell2019-aq}
\item[A7]\hypertarget{A7}{Data Cards} (\hyperlink{T7}{T7}): \citet{Pushkarna2022-sp}
\item[A8]\hypertarget{A8}{Healthsheet} (\hyperlink{T8}{T8}): \citet{Rostamzadeh2022-ie}
\item[A9]\hypertarget{A9}{Value Cards} (\hyperlink{T9}{T9}): \citet{Shen2022-bl}
\item[A10]\hypertarget{A10}{Responsible AI License} (\hyperlink{T10}{T10}): \citet{Contractor2022-he}
\item[A11]\hypertarget{A11}{Algorithmic Equity Toolkit} (\hyperlink{T11}{T11}): \citet{Krafft2021-no}
\item[A12]\hypertarget{A12}{Algorithmic Impact Assessments} (\hyperlink{T12}{T12}): \citet{Reisman2018-df}
\item[A13]\hypertarget{A13}{Fairlearn} (\hyperlink{T13}{T13}): \citet{Agarwal2018-ka}
\item[A14]\hypertarget{A14}{AI Fairness 360} (\hyperlink{T14}{T14}): \citet{Bellamy2019-ho}
\item[A15]\hypertarget{A15}{Interactive Model Cards} (\hyperlink{T15}{T15}): \citet{Crisan2022-zz}
\item[A16]\hypertarget{A16}{SageMaker Clarify} (\hyperlink{T16}{T16}): \citet{Hardt2021-rv}
\item[A17]\hypertarget{A17}{InterpretML} (\hyperlink{T17}{T17}): \citet{Nori2019-ow}
\item[A18]\hypertarget{A18}{Aequitas} (\hyperlink{T18}{T18}): \citet{Saleiro2018-do}
\item[A19]\hypertarget{A19}{Lanugage Interpretability Tool} (\hyperlink{T19}{T19}): \citet{Tenney2020-fo}
\item[A20]\hypertarget{A20}{What-If Tool} (\hyperlink{T20}{T20}): \citet{Wexler2020-kf}
\item[A21]\hypertarget{A21}{Model Card Authoring Toolkit} (\hyperlink{T21}{T21}): \citet{Shen2022-bl}
\item[A22]\hypertarget{A22}{LiFT} (\hyperlink{T22}{T22}): \citet{Vasudevan2020-mm}
\item[A23]\hypertarget{A23}{Data Sheets} (\hyperlink{T4}{T4}): \citet{Boyd2021-dr}
\item[A24]\hypertarget{A24}{Consequence Scanning} (\hyperlink{T27}{T27}): \citet{Smiee2022-bq}
\item[A25]\hypertarget{A25}{Model Cards} (\hyperlink{T6}{T6}): \citet{Nunes2022-jl}
\item[A26]\hypertarget{A26}{Fairlearn} (\hyperlink{T13}{T13}): \citet{Bird2020-lu}
\item[A27]\hypertarget{A27}{AI Fairness 360, Fairlearn} (\hyperlink{T14}{T14}, \hyperlink{T13}{T13}): \citet{Deng2022-rj}
\item[A28]\hypertarget{A28}{Aequitas} (\hyperlink{T18}{T18}): \citet{Richardson2021-kp}
\item[A29]\hypertarget{A29}{Aequitas, AI Fairness 360, Fairlearn, What-If Tool} (\hyperlink{T18}{T18}, \hyperlink{T14}{T14}, \hyperlink{T13}{T13}, \hyperlink{T20}{T20}): \citet{Lee2021-fn}
\item[A30]\hypertarget{A30}{AI Fairness 360, What-If Tool} (\hyperlink{T14}{T14}, \hyperlink{T20}{T20}): \citet{Gu2021-bl}
\item[A31]\hypertarget{A31}{InterpretML} (\hyperlink{T17}{T17}): \citet{Kaur2020-ht}
\item[A32]\hypertarget{A32}{audit-AI} (\hyperlink{T24}{T24}): \citet{audit_ai-uh}
\item[A33]\hypertarget{A33}{Ethical Compass} (\hyperlink{T25}{T25}): \citet{IDEO_compass-bg}
\item[A34]\hypertarget{A34}{Judgment Call} (\hyperlink{T26}{T26}): \citet{Ballard2019-nw}
\item[A35]\hypertarget{A35}{Algorithmic Equity Toolkit} (\hyperlink{T11}{T11}): \citet{Katell2020-ia}
\item[A36]\hypertarget{A36}{Consequence Scanning} (\hyperlink{T27}{T27}): \citet{Brown2019-ug}
\item[A37]\hypertarget{A37}{Algorithmic Accountability Policy Toolkit} (\hyperlink{T23}{T23}): \citet{AI_Now_Institute2018-hu}
\end{enumerate}

\section{Summary of themes and codes}
\label{app:thematicanalysis}

\onecolumn
\begin{longtable}{@{}p{0.1\linewidth}p{0.2\linewidth}p{0.65\linewidth}@{}}
\toprule
Theme &
Codes &
Example segment\\* 
\midrule
\endhead
\bottomrule
\endfoot
\endlastfoot
Research contribution       & The tool is the contribution           & "This paper contributes one new professional practice—called data statements—which we argue will bring about improvements in engineering and scientific outcomes while also enabling more ethically responsive NLP technology" [\hyperlink{A2}{A2}]\\
                            & The evaluation is the contribution     & "the core contribution of this paper is the identifcation of gaps between the capabilities of existing open source fairness toolkits and the requirements of practitioners, highlighting the implications of these gaps to inform the development of fairness-related tooling" [\hyperlink{A29}{A29}]\\
                            & The design process is the contribution & "We present the Algorithmic Equity Toolkit project as a case study of the value of participatory design and action research toward greater fairness, accountability, and transparency in algorithmic systems" [\hyperlink{A35}{A35}]\\
Evaluation focus            & User-centered framework                & "Our  results  highlight  challenges  for  designing  interpretability  tools  for  data  scientists,  and,  in  line  with  prior  work  {[}49,  62{]},  we  advocate for similar user-centric evaluations to be conducted  for all stakeholders of interpretability tools and ML models." [\hyperlink{A31}{A31}]\\
                            & Usability metrics                      & "As expected, the most novel feature of the tool, the causal graph view, received the most dwell time of all of the tools (M: 394.5s, Std: 797.9s, total session: 1200s) and the highest number of interaction events (M:11.0:, Std: 5.69)" [\hyperlink{A30}{A30}]\\
                            & Impact on development                  & "The impact of fair ML toolkits in this study is evident when comparing willingness to deploy models before and after fairness analysis" [\hyperlink{A28}{A28}]\\
                            & Inform ongoing development             & "Over the course of 15 months, we conducted internal and external studies and designed WIT based on the feedback we received" [\hyperlink{A20}{A20}]\\
Tool adoption requirements  & Fast paced environment                 & "rapidly on-boarding toolkits due to workplace time constraints" [\hyperlink{A27}{A27}]\\
                            & Easy adoption                          & "we need to enable users to try out the feature without reading documentation" [\hyperlink{A16}{A16}]\\
                            & Resource constraints                   & "Future work will explore further strategies for reducing and re-distributing the work of model documentation" [\hyperlink{A15}{A15}]\\
                            & Validity of findings                   & "Future research is needed to replicate and validate our methods and findings in other decision-making contexts." [\hyperlink{A9}{A9}]\\
Nature of RAI work          & Community stakeholders                 & "we introduce a novel deliberation driven toolkit – the Model Card Authoring Toolkit – to help community members understand, navigate and negotiate a spectrum of models via deliberation and try to pick the ones that best align with their collective values" [\hyperlink{A21}{A21}]\\
                            & Collaborative and interdisciplinary    & "when tackling complex, multi-faceted fairness issues in real-world settings, fairness toolkits need to support interactions and collaborations across diverse roles, including non-engineers" [\hyperlink{A27}{A27}]\\
                            & Complexity                             & "Fairness  is  a  complex  construct  that  cannot  be  captured with  a  one-size-fits-all  solution" [\hyperlink{A14}{A14}]\\
                            & Different forms of expertise           & "When creating datasets that relate to people, and hence their accompanying datasheets, it may be necessary for dataset creators to work with experts in other domains such as anthropology, sociology, and science and technology studies" [\hyperlink{A4}{A4}]\\
                            & Reflective                             & "a view of ethics not only as a procedure, with a list of principles to considered at every turn, but also as a continuous deliberative process about actions being taken and decisions being made" [\hyperlink{A25}{A25}]\\
                            & Across the AI development lifecycle    & "Broaden the scope of fairness toolkits to across the ML development lifecycle" [\hyperlink{A27}{A27}]\\
                            & Across sociopolitical contexts         & "The framework presented here is intended to be general enough to be applicable across different institutions, contexts, and stakeholders" [\hyperlink{A6}{A6}]\\
Values in evaluation design & Representative participants            & "We note that the groups considered in this study do not cover an entire spectrum of the real-world diversity" [\hyperlink{A3}{A3}]\\
                            & Realistic tasks                        & "The selected strategy was to compare how a group of competent ML application developers used different representations in a realistic case, inquiring about their thoughts, choices, and decisions." [\hyperlink{A25}{A25}]\\
                            & Realistic experimental setting         & "Our work has several limitations. First, although we tried to  put data scientists in a realistic setting via Jupyter notebooks  in  our  contextual  inquiry  and  via  visualizations  and  the  results  of  common  exploration  commands  in  our  survey,  we  cannot be certain that this was sufficient" [\hyperlink{A31}{A31}]\\
Evaluation design           & Single user-tool task design           & "They were asked to explore this model card and think out loud about what information they would want to gather, whether it was available, and how they would wish to see this information" [\hyperlink{A15}{A15}]\\
                            & Learn in isolation                     & "we evaluated the setup of the proof-of-concept application with each participant’s model and data. Here, we provided participants with a set of instructions and evaluated the technical suitability and compatibility of model and examples" [\hyperlink{A20}{A20}]\\
                            & Group based activities                 & "Through this focus group, we were able to arrive at a working definition and values of transparency relevant to domains within AI product life cycles" [\hyperlink{A7}{A7}]\\
                            & Workplace recruitment                  & "Participants were recruited through internal communication channels and word-of-mouth" [\hyperlink{A28}{A28}]\\
                            & Personal network-based recruitment     & "Participants were recruited through the Slack channel for an ML Meet-up group the author attends" [\hyperlink{A23}{A23}]\\
                            & Student recruitment                    & "We recruited and explored the perspectives of people who we believe are representative of the future data workforce" [\hyperlink{A15}{A15}]\\
Evaluation activities       & Use of scenarios/public datasets       & "We then “tested” these questions by creating example datasheets for two widely used datasets: Labeled Faces in the Wild and Pang and Lee’s polarity dataset" [\hyperlink{A4}{A4}]\\
                            & Follow up survey                       & "To validate our findings from the interviews with a larger sample size and a broader practitioner population, we conducted an anonymous online survey" [\hyperlink{A29}{A29}]\\
                            & Impact of COVID-19                     & "Due to the COVID-19 crisis, this pilot took place over a Zoom meeting during a regular convening of the coalition" [\hyperlink{A11}{A11}]\\
                            & Interviews                             & "We conducted 21 semi-structured interviews with experts with a wide range of applied, industrial and academic expertise" [\hyperlink{A8}{A8}]\\
Future work                 & Study use in practice                  & "This suggests opportunities for future research to test the use of the physical cards in real-world interactions" [\hyperlink{A9}{A9}]\\
                            & Study use in team environments         & "Future work is needed to explore how teams in organizations collectively use fairness toolkits on real-world tasks. This could enable insight into the ways team dynamics might add additional frictions to toolkit use" [\hyperlink{A27}{A27}]\\
                            & Study long term effect                 & "longitudinal studies might reveal different or more nuanced patterns  of behavior than either our contextual inquiry or our survey" [\hyperlink{A31}{A31}]\\
                            & Consider affected communities          & "Future studies should develop and test best practices for writing effective, thorough Datasheets that consider the lived realities of affected communities" [\hyperlink{A23}{A23}]\\* 
\bottomrule
\end{longtable}
\twocolumn

\end{document}